\documentclass[hyper, a4paper, 11pt, oneside]{JHEP3} 

\usepackage{amssymb,amsmath}
\usepackage{graphicx,epsfig}
\usepackage{epsfig,multicol,bbm}

\newcommand\fverb{\setbox\fverbbox=\hbox\bgroup\verb}
\newcommand\fverbdo{\egroup\medskip\noindent%
			\fbox{\unhbox\fverbbox}\ }
\newcommand\fverbit{\egroup\item[\fbox{\unhbox\fverbbox}]}
\newbox\fverbbox

\title{Analysis of Discrete Symmetries in b-Baryon
Decays}
\author{Aqeel Ahmed\\
	National Centre for Physics and Department of Physics,
Quaid-i-Azam University,
Islamabad, 45320, Pakistan\\
	E-mail: \email{aqeel@ncp.edu.pk}, \email{aqeelhmed@gmail.com}}

\date{\today}

\abstract{A study of the decay channels $\Lambda_{b}\to \Lambda V$, where $V$
is a vector meson $(1^-)$, has been done by putting together kinematical
and dynamical analysis. An intensive use of the helicity formalism is
involved on the kinematical side, while on the dynamical side, Heavy Quark
Effective Theory (HQET) is applied to calculate the hadronic matrix elements
between the baryons $\Lambda_{b}$ and $\Lambda$. The branching ratios($%
\mathcal{BR}$) and helicity asymmetry parameters ($\alpha_{AS}$) for $%
\Lambda_{b}\to \Lambda J/\psi$, $\Lambda_{b}\to \Lambda \rho^0$ and $%
\Lambda_{b}\to \Lambda \omega$ have been calculated. Since both the decay
products are polarized, so they offer interesting opportunities to perform
tests of time reversal, CP violation and of CPT invariance. A model
independent parametrization is done via spin density matrix of the angular
distribution, polarizations and some of polarization correlations of the
decay products. The transverse component of the polarization and two
polarization correlations are sensitive to time reversal violations.
Moreover several CP- and CPT-odd observables are pointed out.}

\keywords{Baryon Decays, Flavor Physics, CP violation, TR violation, CPT, HQET}

\begin{document}

\section{Introduction}

The quest of man to understand the nature and the phenomena occurring in the
universe is as old as the human history. His curiosity led him to modern
sciences to abstract the answer but he is still in search to answer the
unanswered questions in the nature. Particle physics is the study of the
fundamental constituents of matter and the forces governing them. It
endeavors to answer the questions: What are the fundamental constituents of
matter? How they interact? What is the nature of these interactions
(forces)? How these constituent particles are different from each other? Why
the interactions are different on different scales? Can these be unified?
And several other questions essential for our understanding of the Universe.

It is now well established that the fundamental constituent particles are
quarks and leptons. Quarks are of six flavors namely: up($u$), down($d$ ),
strange($s$), charmed($c$), bottom($b$) and top($t$) with different quantum
numbers associated with them. Leptons are also of six flavors, to be exact:
electron($e$), muon($\mu $), tau($\tau $), electron neutrino($\nu _{e} $),
muon neutrino($\nu _{\mu }$) and tau neutrino($\nu _{\tau }$), having
different physical properties. All quarks are fractionally charged while
leptons are integrally charged except neutrinos. All fundamental particles
have their anti-particles with opposite quantum numbers.

These elementary particles experience the four fundamental forces of nature.
These forces are the electromagnetic, strong nuclear, weak nuclear, and
gravity. Electromagnetic force occurs via exchange of a photon and is
experienced by charged particles. Strong nuclear force occurs by the
exchange of gluons and is accountable for the stability of nuclei. Weak
nuclear force is mediated by three particles known as $W^{\ +}$, $W^{\ -}$,
and $Z^{\ 0}$ and is responsible for the radioactive decays. Gravity is
presumably mediated by a graviton and is experienced by massive particles.
All of these force mediators are bosons having integer intrinsic spin.
Electromagnetic and gravity have infinite range because their propagators
are massless while strong and weak forces are short range as their mediators
are massive. The relative strengths of strong nuclear force,
electromagnetic, weak nuclear force and gravity are in the order of $%
1:10^{-2}:10^{-7}:10^{-40}$ respectively. The universality of these
interactions implies that they are gauge forces.

All of the constituent particles of matter and forces governing them are put
in a nutshell known as `Standard Model' and it is the only experimentally
tested model so far. The Standard Model classify all quarks and lepton into
three generations. The first generation
\begin{equation*}
\binom{u}{d},\ \binom{\nu_{e}}{e}
\end{equation*}
is pertinent for the visible universe and the life on earth. The second and
third generations
\begin{equation*}
\binom{c}{s},\ \binom{\nu_{\mu }}{\mu }
\end{equation*}
and
\begin{equation*}
\binom{t}{b},\ \binom{\nu_{\tau }}{\tau }
\end{equation*}
do not exist naturally but can be created either in laboratory or in cosmic
rays by a collision of particles of first generation.

Standard Model satisfactorily explains most of the observable phenomena in
elementary particle physics. Gravity, being incredibly weak compared to the
other forces is not important while studying microscopic particles and is
not elucidated by the standard model. Standard Model incorporates three
forces of nature including\ strong nuclear force, electromagnetic and weak
nuclear force. It is generally believed that the Standard Model will be a
part of final theory which unifies all the forces, known as Theory of
Everything, which combines standard Model with Gravity.

One of the area of standard model, which is still poorly tested
experimentally and has the potential to provide indications of new physics,
is the physics of weak decays of heavy hadrons. The heavy hadrons contain
quarks $c$ and $b$ and other lighter quarks $u$, $d$ and $s$, form lighter
baryons. The $b-$baryons is a heavy hadron because it contains $b$-quark and
has a mass of $5.28 \ GeV/c^{2} $, which is more than five times the mass of
the proton. The $b-$baryon would be stable if $b-$quark and companion
anti-quark doesn't have weak charge. Because they do, and because this
hadron is heavier than many other hadrons, there are many channels in which
it can decay. All of these decays involve the b-quark transforming itself
into another lighter quark, which could be a $c$, $s$, $u$ or $d$ quark.

Now the question arises: why we study b-hadron decays? The most obvious
reason is that the $b$-hadrons are the heaviest hadrons, as the top quark
decays before it can hadronize. The fact that $b$-hadron is heavy has two
important consequences: $b$-hadrons decays show an extremely rich
phenomenology and theoretical techniques using an expansion in the heavy
mass allow for model-independent predictions. The large available phase
space and the the possibility for large CP-violating asymmetries in the $b$%
-hadron decays make it a topic of rich phenomenological study. The large
CP-violating feature of $b$-hadrons is in contrast to the Standard Model
expectations for the decays of $K$ and $D$ mesons. The pattern of CP
violation in $K$ and $B$ system just represents the hierarchy of the CKM
matrix. The $b$-hadron system offers an excellent laboratory to
quantitatively test the CP-violating and time reversal violating sector of
the Standard Model, determine fundamental parameters, study the interplay of
strong and electroweak interactions and some of search for New Physics (NP).

In this thesis we have worked on hadronic decays of $\Lambda_{b}$ baryon
because a huge statistics of beauty hadrons are expected to be produced at
the CERN-LHC proton-proton collider started last year. Obviously this will
lead to a thorough study of discrete symmetries, C, P and T in b-quark
physics, in the framework of the Standard Model (SM) as well as beyond the
Standard Model. It is also well known that the violation of CP symmetry via
the Cabibbo-Kobayashi-Maskawa (CKM) mechanism is one of the cornerstone of
the Standard Model of particle physics in the electroweak sector. In LHCb
experiment non-leptonic and leptonic b-baryon decays may allow us to get
information about the CKM matrix elements, analysis of the C-P-T operators
may be performed and different non-perturbative aspects of QCD may also be
investigated.

Looking for CP and Time Reversal (TR) violation effects in b-baryon decays
can provide us a new field of research. Firstly, TR violation can be seen as
a complementary test of CP violation by assuming the correctness of the CPT
theorem. Secondly, this can also be a path to follow in order to search for
processes beyond the Standard Model. Various observables which are T-odd
under time reversal operations can be measured, so that $\Lambda_{b}$-decay
seems to be one of the most promising channel to reveal TR violation signal.

Although some CP violation and also a direct TR violation $\cite{TRV}$ have
been detected experimentally, the nature of such symmetry violations has not
yet been clarified so far. More precisely, the prediction of the size of the
violation in some weak decays is strongly model dependent, which stimulates
people to search for signals of new physics (NP)$\cite{TMA},\cite{datta1},%
\cite{datta2},\cite{Pakvasa}$, beyond the standard model (SM). For example,
the decays involving the transition $b\to s$ present CPV parameters, like
the $B^{0}-\bar{B}^{0}$ mixing phase$\cite{IBD},\cite{bona}$ and the
transverse polarization of spinning decay products of $\Lambda_{b}$ $\cite%
{TMA}$, which are very small in SM predictions, but are considerably
enhanced in other models. In particular, recent signals of NP have been
claimed in B decays: the CP violating phases of $B\to K\pi$ $\cite{IBD}$ and
$B\to \phi J/\psi$ $\cite{bona}$ may be considerably greater than predicted
by SM. Also $\Lambda_{b}$ decays $\cite{TMA},\cite{mi},\cite{ACL}$ are
suggested as new sources of CPV and TRV parameters, especially in view of
the abundant production of this resonance in the forthcoming LHC accelerator.

This thesis is organized as follows: in chapters 2 and 3 we fill our toolbox
with the necessary ingredients. After giving a bird eye view of the Standard
Model and Discrete Symmetries in chapter 2, we present some basic tools in
chapter 3, like operator product expansion, Heavy Quark Effective Theories
and QCD factorization to analyze our decay. A discussion of the effective $%
\Lambda_{b}\rightarrow \Lambda V$ Hamiltonian and evolution of Form Factors
are also the subjects of chapter 3.

In chapter 4 we worked out some model independent tests of TRV, CPV and CPT
invariance in hadronic $\Lambda_{b}$ decays of the type $\Lambda_{b}\to
\Lambda V$, where V denoting a $J^{P} = 1^{-}$, resonance, either the $%
J/\psi $ or a light vector meson, like $\rho$, $\omega$. Each resonance
decays, in turn, to more stable particles, like, e. g.,$\Lambda\to p\pi$ and
$J/\psi\to l^{+}l^{-}$. We parameterized, by means of the spin density matrix
(SDM), the angular distribution and the polarizations of the decay products,
without introducing any dynamic assumption at all. Then we study the
behavior of these observables under CP and T, singling out those which are
sensitive to T, CP and CPT violations. Our approach resembles the one
proposed by Lee and Yang $\cite{ly}$ and by Gatto $\cite{gatto}$ many years
ago, to use hyperon decays for the same tests.

We derive the expressions of the spin density matrices, angular distribution
and polarizations of the decay products in the above mentioned decays by
using the Jacob-Wick-Jackson Helicity formalism. We also present a
parametrization of the angular distribution and polarizations and point out
tests for TRV, CPV and CPT. In the last chapter we have put the numerical
analysis of our work and draw the conclusions.

\section{Standard Model and Discrete Symmetries}

The Standard Model (SM) of particle physics represents our understanding of
the fundamental nature of the universe. It describes the basic constituents
of which all matter is made of, which are three families of quarks and
leptons, and the forces. The wish for a simple and consistent description of
all observed phenomena has lead to a quantum field theory that successfully
describes many physical observables and is consistent with the Standard
Model. It is a gauge quantum field theory.

Although Standard Model satisfactorily describes most of the observed
phenomena of elementary Particle Physics, several questions, important for
our understanding of the Universe, remain unanswered. These problems are
often entitled as Physics Beyond Standard Model such as the hierarchy
problem, the missing matter problem (dark matter and dark energy),
phenomenon of generation, Time reversal and CP-violation and Baryogenesis.

\subsection{Introduction to the Standard Model}

Strong and electroweak interactions between elementary particles are best
described by the Standard Model. Standard Model lagrangian is made of two
parts: one for electro-weak interactions known as the Glashow-Weinberg-Salam
model $\cite{GWS}$, and the other is the Quantum Chromodynamics (QCD) for
the strong interactions $\cite{MW}$. It unifies all known experimental data
concerning particle interactions via the gauge group $SU_{C}(3)\otimes
SU_{L}(2)\otimes U(1)$. The gauge fields of color $SU_{C}(3)$ are
responsible for binding the quarks together, while the gauge fields of $%
SU_{L}(2)\otimes U(1)$ mediate the electromagnetic and weak interactions.
Because of the low mass of the elementary particles, gravity doesn't give
effects comparable to the other forces, so the Standard Model does not
include this interaction.

The symmetries that characterize the Standard Model are: $SU_{L}(2)$ of weak
isospin $I$, $U(1)$ of hypercharge $Y$ and $SU_{C}(3)$ of color $C$. The $%
SU_{L}(2)$ part of the weak interaction gives rise to a triplet of vector
bosons $W$ associated with the quantum number of weak isospin. To the $U(1)$
component contributes one single boson $B$ associated with the weak
hypercharge $Y$, which is a combination of the electric charge $Q$ and the
third component of the weak isospin $I_{3}$,

\begin{equation}
Y=2(Q-I_{3})
\end{equation}

In particular, the part of the theory that describes the electro-weak
interactions has to be invariant under $SU_{L}(2)\otimes U(1)$, while QCD
has the symmetry $SU_{C}(3)$. Altogether, there are nineteen free parameters
in the theory, suggesting it is not a complete account of particle
interactions. There are three coupling constants for the groups in $%
SU_{C}(3)\otimes SU_{L}(2)\otimes U(1)$, two parameters in the Higgs sector,
6 quark masses. 3 mixing angles and one phase, 3 lepton masses, and the QCD
vacuum angle.

\subsection{Quantum Chromodynamics}

Quantum chromodynamics (QCD) is a non-abelian gauge theory for strong
interactions. Quark interactions can be described using a new quantum
number: the color. In particular each quark can have three different colors,
which generate the group $SU(3)_{color}$. Leptons do not carry color that is
the reason why they do not experience strong interactions. Hadrons are bound
states of quarks or quark and anti quark. Known hadrons are color singlets $%
i.e.$ Color is confined in a hadron.

$q_{a}:$ belong to fundamental representation of $SU_{c}(3)$

\begin{equation}
q_{a}\rightarrow q_{a}^{\prime }=U_{a}^{b}q_{b}
\end{equation}
when there is more than one type of states, e.g. $q_{a}\ (a=1,2,3)$ and
there exists transformations $SU_{c}(3)$ between the different states, with

\begin{equation}
U(x)=e^{\frac{i}{2}\lambda _{A}\Lambda _{A}(x)}  \label{qc1}
\end{equation}
where $\lambda _{A}$ are Gell-Mann matrices , $A=1,...,8$
\begin{equation}
UU^{\dag }=1
\end{equation}
\begin{equation}
\det U=1
\end{equation}
Here $q_{a}$ for a particular quark flavor $q$ form the fundamental
representation of the color $SU(3)$ group and $\lambda _{A}$ are the eight
matrix generators of the group $SU_{c}(3).$

Quarks are spin 1/2 particles. The lagrangian density for free quarks is,
\begin{equation}
L=\bar{q}^{a}i\gamma ^{\mu }\partial _{\mu }q_{a}-\bar{q}^{a}mq_{a}
\end{equation}
where
\begin{equation*}
q_{a}=\left(
\begin{array}{c}
u_{a} \\
d_{a} \\
s_{a}%
\end{array}
\right)\text{ \ \ \ \ \ \ \ and \ \ \ \ \ \ \ } m=\left(
\begin{array}{ccc}
m_{u} & 0 & 0 \\
0 & m_{d} & 0 \\
0 & 0 & m_{s}%
\end{array}
\right)
\end{equation*}
is clearly invariant under the SU(3) transforation with $\Lambda $\
constant. For the local gauge transformation Eq $\eqref{qc1}$, with $\Lambda
(x)$\ as a function of space-time, we must replace $\partial _{\mu }$\ by
its covariant derivative $D_{\mu}$:

\begin{equation}
D_{\mu }=(\partial _{\mu }-\frac{i}{2}g_{s}\lambda .G_{\mu })=(\partial
_{\mu }-\frac{i}{2}g_{s}\lambda _{A}G_{A\mu })
\end{equation}%
where $g_{s}$\ is a scale parameter, the coupling constant and $G_{A\mu }$\
are vector gauge fields, their number being equal to the generators of $%
SU_{c}(3)$ group i.e. 8. Now the Lagrangian density is given by:

\begin{equation}
L=\bar{q}^{a}i\gamma ^{\mu }(\partial _{\mu }-\frac{i}{2}g_{s}\lambda
_{A}G_{A\mu })_{a}^{b}q_{b}-\bar{q}^{a}mq_{a}-\frac{1}{4}G_{A}^{\mu \nu
}G_{A\mu \nu }
\end{equation}

The eight gauge vectors bosons $G_{A\mu }$\ are called gluons. They are
mediators of strong interaction between quarks just as photons are mediators
of electromagnetic force between electrically charged particles. The gauge
transformation given in Eq$\ (24)$ is called the non-abelian gauge
transformation. As non-Abelian gauge transformation was first considered by
Yang Mills and gauge bosons are sometimes called Yang-Mills Fields.

\subsection{Spontaneous Symmetry Breaking}

The Higgs field which is associated with the Higgs particle interacts with
the quarks, leptons, and weak bosons to give them masses. The coupling of
the the Higgs is the only thing that differentiates the three generations of
quarks and leptons. The Higgs is the only particle in the Standard Model
which has not yet been observed experimentally.

The striking inconsistency of the masses of the gauge bosons with gauge
invariance seeks for a satisfying mechanism to explain these properties
observed in experiments. In the SM, this is provided by the so-called Higgs
mechanism. The Higgs mechanism is a theory which explains the masses of
particles. The particles acquire mass as they move through the Higgs field.
This is an essential part of the standard model as without it, the theory
suggests all particles would be massless (For more details on can see for
example $\cite{FR}$). To prove this mechanism, experiments are trying to
detect the Higgs boson a quantum of the Higgs field. In this model, the mass
is generated by the interaction of particles with the Higgs complex scalar
field $\phi $,
\begin{equation}
\phi =(\phi ^{+},\phi ^{0})
\end{equation}
To illustrate the idea consider a $U(1)$ group and a complex scalar field:
\begin{equation}
\mathcal{L}_{Higgs}=\partial ^{\mu }\bar{\phi}\partial _{\mu }\phi -U(\phi )
\end{equation}
with
\begin{equation}
U(\phi )=\mu ^{2}\bar{\phi}\phi +\lambda(\bar{\phi}\phi )^{2}
\end{equation}
The potential $U\left( \phi \right) $ has rotational symmetry and has its
minimum on a circle at
\begin{equation}
|\phi|^{2}=\frac{-\mu ^{2}}{2\lambda }
\end{equation}
This means that, in principle, any state with
\begin{equation}
|\phi |^{2}=\frac{v^{2}}{2},\text{ with } v^{2}=\frac{-\mu^{2}}{\lambda}
\end{equation}
could be the ground-state in this potential. This is a classical
approximation to the vacuum expectation of $\phi ,$ i.e the ground state,
\begin{equation}
\langle 0|\phi |0\rangle =\sqrt{\frac{-\mu ^{2}}{2\lambda }}\equiv \frac{v}{%
\sqrt{2} }
\end{equation}
breaks the symmetry.
\begin{figure}[h]
\centering
\includegraphics[scale=0.3]{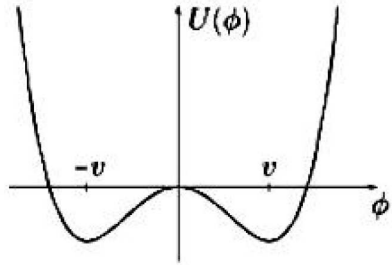}
\caption{The Higgs Potential $U(\protect\phi)$}
\end{figure}
In other words Lagrangian is invariant but the Hamiltonian is not. Regarding
a small excitation of this ground state,
\begin{equation}
\phi =\frac{1}{\sqrt{2}}(v+H+i\eta )
\end{equation}
where
\begin{equation}
\langle0|H|0\rangle =0
\end{equation}
\begin{equation}
U(H)=-\frac{1}{2}\lambda v^{2}[(v+H)^{2}+\eta ^{2}]+ \frac{1}{4}\lambda
[(v+H)^{2}+\eta ^{2}]^{2}
\end{equation}
and putting in the covariant derivative here, yields the Lagrangian density
for Higgs field,
\begin{equation}
\mathcal{L} =\frac{1}{2}(\partial ^{\mu }H)(\partial _{\mu }H)+\frac{1}{2}
(\partial ^{\mu }\eta )(\partial _{\mu }\eta )-\frac{1}{2}\lambda
v^{2}[(v+H)^{2}+\eta ^{2}]+ \frac{1}{4}\lambda[(v+H)^{2}+\eta ^{2}]^{2}
\end{equation}
\begin{equation*}
m_{H}^{2}=\lambda v ^{2},\ m_{\eta }^{2}=0.
\end{equation*}
Lagrangian is invariant under global gauge transformation
\begin{equation}
\phi ^{\prime }(x)=U^{-1}\phi(x) U=e^{i\Lambda }\phi (x)
\end{equation}
but not under the local gauge transformation when $U$ is a function of $x$.
Gauge invariance requires a vector field $B_{\mu }$
\begin{equation}
B_{\mu }\rightarrow B_{\mu }-\frac{1}{g}\partial _{\mu }\Lambda
\end{equation}
Gauge invariant Lagrangian can be obtained by replacing $\partial _{\mu }$
with $D_{\mu}$,
\begin{equation}
\partial _{\mu }\rightarrow D_{\mu} = \partial _{\mu }-igB_{\mu }
\end{equation}
\begin{equation}
\mathcal{L} =-\frac{1}{4}B^{\mu \nu }B_{\mu \nu }+(\partial _{\mu }+igB_{\mu
})\ \bar{\phi}\ (\partial _{\mu }-igB_{\mu })\ \phi -U(\phi )
\end{equation}
where $B_{\mu\nu}=\partial _{\mu }B_{\nu}-\partial _{\nu }B_{\mu}$. The
unwanted zero mass mode due to spontaneous symmetry breaking can be
eliminated by means of field dependent gauge transformation
\begin{equation}
\phi (x)\rightarrow \frac{1}{\sqrt{2}}[v+H(x)]\ e^{i\frac{\eta (x)}{v}}
\end{equation}
\begin{equation}
B_{\mu }\rightarrow B_{\mu }-\frac{1}{vg}\partial _{\mu }\eta (x)
\end{equation}
\begin{equation}
\mathcal{L}=-\frac{1}{4}B^{\mu \nu }B_{\mu \nu }+\frac{1}{2}(\partial ^{\mu
}H)(\partial _{\mu }H)+\frac{1}{2}g^{2}(v^{2}+2vH+H^{2})\ B^{\mu }B_{\mu }+%
\frac{1}{2}v^{2}\lambda (H+v)^{2}-\frac{\lambda }{4}(H+v)^{4}
\end{equation}
\begin{equation*}
m_{B}^{2}=\frac{1}{2}g^{2}v^{2},\ m_{H}^{2}=v^{2}\lambda
\end{equation*}
The vector boson becomes massive, the Goldstone field $\eta (x)$ has been
transformed away, it has been eaten away by $B_{\mu }$ to give it a
longitudinal component.

\subsection{Cabibo-Kobayashi-Masakawa Matrix (CKM )}

The mechanism of quark mixing is a fundamental pillar of the Standard Model.
This formalism successfully describes transitions between the quark
families. It makes use of the Cabibbo-Kobayashi-Maskawa Matrix, a $3\times 3$
unitary matrix, which can be parameterized by four independent parameters. A
precise determination of the Standard Model parameters allows to test
predictions derived from these input numbers.

The Higgs couplings not only give masses to the quarks and leptons, but also
allows transitions between generations since the mass eigenstates are not
equal to the weak eigenstates. The advantage of the mass basis is that the
quark states can be identified experimentally by their masses. Because the
weak couplings of the three generations are all identical, linear
combinations of the three weak eigenstates can be constructed so that the up
type quarks ($u$,$c$ and $t$) are both weak and mass eigenstates. The weak
eigenstates of the down type quarks are denoted $d^{\prime },s^{\prime }$
and $b^{\prime }$ and the mass eigenstates are denoted $d,s$ and $b$.

These two bases are related by the Cabibbo-Kobayashi-Maskawa (CKM) matrix,

\bigskip

\begin{equation}
\left(
\begin{array}{c}
d^{\prime} \\
s^{\prime} \\
b^{\prime}%
\end{array}
\right) =\left(
\begin{array}{ccc}
V_{ud} & V_{us} & V_{ub} \\
V_{cd} & V_{cs} & V_{cb} \\
V_{td} & V_{ts} & V_{tb}%
\end{array}
\right) \left(
\begin{array}{c}
d \\
s \\
b%
\end{array}
\right)
\end{equation}
Thus the weak eigenstates can be written in terms of the mass eigenstates as

\begin{equation}
|b^{\prime}\rangle =V_{td}|d\rangle +V_{ts}|s\rangle +V_{tb}|b\rangle
\end{equation}
In the lepton sector there is an analogous matrix called
Maki-Nakagawa-Sakata (MNS) matrix.

The weak interaction only produces transitions between the weak eigenstates
of the same generation:($u\rightarrow d^{\prime}$), ($c\rightarrow
s^{\prime} $), and ($t\rightarrow b^{\prime} $). The mass eigenstates are
the states observed experimentally, because the quarks are identified by
their masses. Because $|V_{tb}|$ is much greater than $|V_{ts}|\ $and $%
|V_{td}|\ $, the bottom quark mass eigenstate is mostly b$^{\prime }$ and
very little s$^{\prime }$ and d$^{\prime }$. Therefore the decays of the top
quark to the bottom quark (emitting a $W^{\ +}$) are much more common than
decays of the top quark to the strange and down quarks. So, the probability
of each transition is governed by the overlaps of the mass and weak
eigenstates which is described by the CKM matrix.

\subsubsection{Parametrization of CKM Matrix}

With three generations of fundamental fermions, the CKM matrix can be
parameterized with four parameters: three Euler angles and one phase.
Wolfenstein $\cite{pckm}$ noticed that $|V_{cb}|^{2}\simeq |V_{us}|^{2}$ and
proposed to use $|V_{us}|=\lambda \simeq 0.22$ as an expansion parameter for
the elements of the CKM matrix after the experimental observation that the b
quark decays predominantly to the charm($|V_{cb}|>>|V_{ub}|$). Because the
CKM matrix is a basis transformation, it must be unitary. This constraint
reduces the number of free parameters in the CKM matrix. Without changing
the Lagrangian and hence any observables, the phases of quarks in the
standard Model Lagrangian can be changed. This can be used to remove another
five free parameters from the CKM matrix (these are the five relative phases
of the quark fields). The Wolfenstein parametrization $\cite{Wol}$ of the
CKM matrix exploits the smallness of the off-diagonal elements to construct
a representation in which the relationships between the elements are
manifest:
\begin{equation}
V_{CKM}=\left(
\begin{array}{ccc}
1-\frac{\lambda^{2}}{2}-\frac{\lambda^{4}}{8} & \lambda & A\lambda
^{3}(\rho-i \eta ) \\
-\lambda+\frac{1}{2}A^{2}\lambda^{5}(1-2(\rho+i\eta)) & 1-\frac{\lambda ^{2}%
}{2}-\frac{1}{8}\lambda^{4}(1+4A^{2}) & A\lambda ^{2} \\
A\lambda ^{3}(1-\bar{\rho} -i \bar{\eta} ) & -A\lambda ^{2}+\frac{1}{2}%
A\lambda^{4}(1-2(\rho+i\eta)) & 1-\frac{A^{2}\lambda^{4}}{2}%
\end{array}
\right)  \label{pckm}
\end{equation}
where
\begin{equation}
\bar{\rho}=\rho\left(1-\frac{\lambda^{2}}{2} \right) \text{ and } \bar{\eta}%
=\eta\left(1-\frac{\lambda^{2}}{2} \right)
\end{equation}
and $\eta$ plays the well-known role of the CP-violating phase in the
Standard Model framework. From the CKM matrix, expressed in terms of the
Wolfenstein parameters and constrained with several experimental data, we
will take in our numerical applications,
\begin{equation*}
0.076< \rho< 0.380 \text{ and } 0.280<\eta< 0.455.
\end{equation*}
The values for $A$ and $\lambda$ are assumed to be well determined
experimentally:
\begin{equation*}
\lambda=0.2265\text{ and } A=0.801
\end{equation*}

\subsection{Discrete Symmetries}

Most of the symmetries in elementary-particle physics are continuous. A
typical example is the symmetry generated by rotations around an axis, where
the angle of rotation can assume any value between zero and 2$\pi$. In
addition to continuous symmetries, there are also discrete symmetries, for
which the possible states assume discrete values classified with the help of
a few integers. For instance, snowflakes exhibit the discrete symmetry of
rotations under $60^{\circ}$. and crystals exhibit various types of discrete
symmetries. In elementary-particle physics there are three discrete
symmetries of basic importance: parity, charge conjugation and time-reversal.

Parity is the reflection of space coordinates and will be denoted by $%
\mathbf{P}$. Under parity there are two states, the object and its space
reflection. Parity is familiar from quantum mechanics, where the eigenstates
of Hamiltonian are classified according to their properties under space
reflection. For spherically symmetric potentials the wave functions are
proportional to the spherical harmonics $Y_{m}^{l}(\theta,\phi)$ whose
parity is $(-1)^{l}$. For a long time it was assumed that the fundamental
interactions respect $\mathbf{P}$, but in 1956 a critical review of
experimental evidence led two theoreticians, T. D. Lee and C. N. Yang, to
suggest that parity may be violated by the weak interactions $\cite{Lee}$.
One year later, an experiment led by C. S. Wu brought the proof that the $%
\mathbf{P}$ symmetry is indeed violated by weak interactions.

The symmetry of charge conjugation, to be denoted by $\mathbf{C}$, exchanges
particles with antiparticles. One can imagine building an antiworld by
replacing all particles by antiparticles. In the antiworld the three
interactions gravity, the strong force, and electromagnetism are the same,
but the weak interactions are different. For example in the antiworld
muon-type antineutrinos are right-handed and produce $\mu^{+}$ which are
also right-handed. In comparison neutrinos are left-handed and always
produce, in high-energy reactions, left-handed $\mu^{-}$. In the weak
interactions the $\mathbf{C}$ symmetry is broken. However, it was assumed,
at that time, that the observed processes do respect the combined $\mathbf{CP%
}$ transformation, the one obtained by applying both $\mathbf{C}$ and $%
\mathbf{P}$ transformations.

There is a fundamental reason why $\mathbf{CP}$ symmetry plays a crucial
role. It is intimately linked to the time-reversal transformation ($\mathbf{T%
}$). This transformation consists of ``looking'\ at an experiment running
backward in time. Although, at the macroscopic level, one can distinguish
the real sequence of events from the time-reversed one in terms of
large-scale phenomena such as entropy or the expansion of the Universe. This
is not a priori evident for microscopic interactions, i.e. it is not a
priori evident that the amplitudes for reactions and for the time-reversed
reactions are equal.

The analysis of $\mathbf{CP}$ violation is facilitated by an important
theorem known as $\mathbf{CPT}$ theorem. It states that any local field
theory based on special relativity and quantum mechanics is invariant under
the combined action of $\mathbf{C}$, $\mathbf{P}$, and $\mathbf{T}$. A
consequence of the theorem is that $\mathbf{CP}$ symmetry implies $\mathbf{T}
$ symmetry, because any CP violation should be compensated by $\mathbf{T}$
violation. Until 1964 the decays and interactions of particles showed that
the $\mathbf{CP}$ symmetry was conserved; this created the belief that
microscopic phenomena also obey the $\mathbf{T}$ symmetry. In 1964 CP
violation was observed in an experiment dedicated to the study of $K^{0}$
and $\bar{K}^{0}$ mesons. Since then it has become an active topic of
research, with $\mathbf{CP}$ violation having been observed so far in the $%
\mathbf{K}$ and the $\mathbf{B}$ mesons.

\section{Physics of Heavy Quarks and Beauty(b)-Hadrons}

In 1964, G. Zweig and M. Gell-Mann independently proposed that hadrons are
made up of constituents, called \emph{quarks} by Gell-Mann. The particle
that experience strong interaction are called \emph{Hadrons}. They are the
bound states of quarks and are color singlets. We can divide the hadrons
into two large classes, \emph{Meson}(integer spin) and \emph{Baryon} (half
integer spin). Mesons are made up of quark-antiquark $(q\bar{q})$ system
where as baryons are made up of quark-quark-quark $(qqq)$ system. This
scheme extends the isospin internal symmetry, which is based on the group $%
SU(2)$ to $SU(3)$, a larger unitary group. We immediately stress that the $%
SU(3)$ symmetry has two very different roles in Particle Physics:

\begin{enumerate}
\item The classification of the hadrons, or rather the hadrons with up (u),
down (d) and strange (s) quarks.

\item The symmetry of the charges of one of the fundamental forces, the
strong force.
\end{enumerate}

For quarks we have representation $3$ and for antiquark we have $\bar{3}$ in
$SU(3)$. Mesons are the members of the multiplets belonging to the product
of $3\otimes\bar{3}$. Group theory tells us that we can write them in
irreducible representation of
\begin{equation}
3\otimes\bar{3}=8\oplus1
\end{equation}
While baryons have classification in $SU(3)$ multiplets is $%
3\otimes3\otimes3$ and their irreducible representations are given by
\begin{equation}
3\otimes3\otimes3=10\oplus8^{\prime}\oplus8\oplus1
\end{equation}

\subsection{Heavy Quark Physics}

For many reasons the strong interactions of hadrons containing heavy quarks
are easier to understand than those of hadrons containing only light quarks.
The first is asymptotic freedom, the fact that the effective coupling
constant of QCD becomes weak in processes with large momentum transfer,
corresponding to interactions at short distance scales. At large distances,
on the other hand, the coupling becomes strong, leading to nonperturbative
phenomena such as the confinement of quarks and gluons on a length scale $%
R_{had}\sim 1/\Lambda _{QCD}\sim 1 fm$, which determines the size of
hadrons. Roughly speaking, $\Lambda _{QCD}\sim 0.2GeV\ $is the energy scale
that separates the regions of large and small coupling constant. When the
mass of a quark $Q$ is much larger than this scale, $m_{Q}\gg \Lambda _{QCD}$%
, it is called a heavy quark. The quarks of the Standard Model fall
naturally into two classes: up, down and strange are light quarks, whereas
charm, bottom and top are heavy quarks. For heavy quarks, the effective
coupling constant $\alpha _{s}(m_{Q})$ is small, implying that the strong
interactions are perturbative and much like the electromagnetic interactions.

Systems composed of a heavy quark and other light constituents are like
Hydrogen atom but it is more complicated because it also involves the
gluonic interactions with quarks and themselves. The size of such systems is
determined by $R_{had}$, and the typical momenta exchanged between the heavy
and light constituents are of order $\Lambda _{QCD}$. The heavy quark is
surrounded by a complicated, strongly interacting cloud of light quarks,
antiquarks and gluons. As in the case of charm and bottom quarks, the masses
are $\sim 1.5$ $GeV $ and $\sim 4.9$ $GeV$, respectively,and $\Lambda _{QCD}$
is $\sim 0.2 GeV$. In such systems the heavy quark is almost on-shell; its
momentum fluctuates around the mass shell by an amount of order $\Lambda
_{QCD}$. The corresponding fluctuations in the velocity of the heavy quark
vanish as $\Lambda _{QCD}/m_{Q}\rightarrow 0$. The velocity becomes a
conserved quantity and is no longer a dynamical degree of freedom. Because
the velocity does not depend on the heavy quark mass, different heavy quarks
interact identically in the heavy quark mass limit. This is known as \emph{%
flavor symmetry}. The heavy quark spin also decouples from the strong
interaction. The decoupling of the spin in the heavy quark limit leads to
the \emph{spin symmetry}.

Therefore, the light degrees of freedom are blind to the flavor (mass) and
spin orientation of the heavy quark. They experience only its color field,
which extends over large distances because of confinement. In the rest frame
of the heavy quark, it is in fact only the electric color field that is
important; relativistic effects such as color magnetism vanish as $%
m_{Q}\rightarrow \infty $. Since the heavy-quark spin participates in
interactions only through such relativistic effects, it decouples. These two
symmetries have important consequences, especially for the decays of beauty
hadrons to lighter hadrons. These symmetries are only true in the heavy
quark limit and are violated at order $\frac{\Lambda _{QCD}}{m_{Q}}$. These
observations can be formalized by writing the Standard Model Lagrangian as
an expansion in $\frac{1}{m_{Q}}$.

Heavy-quark symmetry is an approximate symmetry, and corrections arise since
the quark masses are not infinite. In many respects, it is complementary to
chiral symmetry, which arises in the opposite limit of small quark masses.
There is an important distinction, however, whereas chiral symmetry is a
symmetry of the QCD Lagrangian in the limit of vanishing quark masses,
heavy-quark symmetry is not a symmetry of the Lagrangian (not even an
approximate one), but rather a symmetry of an effective theory that is a
good approximation to QCD in a certain kinematic region. Nevertheless,
results derived on the basis of heavy-quark symmetry are model-independent
consequences of QCD in a well-defined limit. The symmetry-breaking
corrections can be studied in a systematic way. To this end, it is however
necessary to cast the QCD Lagrangian for a heavy quark,
\begin{equation}
\mathcal{L}=\bar{\Psi}_{Q}\left(iD_{\mu }\gamma ^{\mu }-m_{Q}\right)\Psi _{Q}
\end{equation}
into a form suitable for taking the limit $m_{Q}\rightarrow \infty$.

\subsection{Heavy Quark Effective Theory}

The QCD Lagrangian does not explicitly contain heavy quark spin-flavor
symmetry as $m_{Q}\rightarrow \infty $. It is often helpful to use an
effective field theory for QCD in which these symmetries are apparent. The
effective field theory is constructed so that only inverse powers of $m_{Q}$
appear in the effective Lagrangian. The QCD lagrangian describing a quark $Q$
of mass $m_{Q}\ \ $and its interactions with the gluons is given by
\begin{equation}
\mathcal{L}=\bar{\Psi}_{Q}\left(iD_{\mu }\gamma ^{\mu }-m_{Q}\right)\Psi _{Q}
\label{3h1}
\end{equation}
with
\begin{equation*}
D_{\mu }=\partial _{\mu }-ig_{s}T^{a}A_{a\mu }
\end{equation*}
This effective field theory is known as heavy quark effective theory (HQET)
and is describes the dynamics of hadrons containing single heavy quark .

The heavy-quark effective theory (HQET) is constructed to provide a
simplified description of processes where a heavy quark interacts with light
degrees of freedom predominantly by the exchange of soft gluons. At short
distances, i.e. for energy scales larger than the heavy-quark mass, the
physics is perturbative and is described by perturbative QCD. For mass
scales much below the heavy-quark mass, the physics is complicated and
non-perturbative because of confinement. Our goal is to obtain a simplified
description in this region using an effective field theory. To separate
short- and long-distance effects, we introduce a separation scale $\mu $
such that $\Lambda _{QCD}\ll \mu \ll m_{Q}$. The HQET will be constructed in
such a way that it is equivalent to QCD in the long-distance region, i.e.
for scales below $\mu $. In the short-distance region, the effective theory
is incomplete, since some high-momentum modes have been integrated out from
the full theory. The fact that the physics must be independent of the
arbitrary scale, allows us to derive renormalization-group equations, which
can be employed to deal with the short-distance effects in an efficient way.

Compared with most effective theories, in which the degrees of freedom of a
heavy particle are removed completely from the low-energy theory, the HQET
is special in that its purpose is to describe the properties and decays of
hadrons which do contain a heavy quark. In the heavy quark limit ($%
m_{Q}\rightarrow \infty $), the conserved velocity $v _{\mu }$ of the heavy
quark and its four momentum may be decomposed as:

\begin{equation}
p_{\mu}=m_{Q}v _{\mu}+k_{\mu}
\end{equation}
with $v ^{2}=1,$ where $m_{Q}v _{\mu }$ and $k_{\mu }$ are on-shell and
off-shell parts respectively. The components of residual momentum $k$ are
much smaller than $m_{Q}$ and are changed by interactions of the heavy quark
with light degrees of freedom by $\Delta k\sim \Lambda _{QCD}$.

We can separate out the large and small components of the heavy quark field
as
\begin{equation}
h_{v }(x)\equiv e^{i mv \cdot x}\frac{1+\not{v}}{2}\Psi _{Q}(x)
\end{equation}
and
\begin{equation}
H_{v }(x)\equiv e^{i mv \cdot x}\frac{1-\not{v}}{2}\Psi _{Q}(x)
\end{equation}
with the properties $\not{v}h_{v }=h_{v }\ $and $\not {v}H_{v}=-H_{v }$,
respectively. The heavy quark field in terms of the new fields can be
expressed as

\begin{equation}
\Psi _{Q}(x)=e^{-im v \cdot x}(h(x)+H_{v }(x))
\end{equation}
One may split the covariant derivative $D$ into 'longitudinal' and
'transverse' parts as:
\begin{equation}
D_{\bot }=D^{\mu }-v^{\mu }v \cdot D
\end{equation}
with $v \cdot D_{\bot} =0,$ \ \ $\ \{\not{D}_{\bot },\not{v}\}=0,$

Using relations as $\bar{h}_{v }H_{v}=0$ and $\bar{h} _{v }\not{D}_{\bot
}H_{v}=0,$ the lagrangian takes the form

\begin{equation}
\mathcal{L}_{eff}=\bar{h}_{v }i(v \cdot D)h_{v}-\bar{H} _{v }(iv \cdot
D+2m_{Q})H_{v}+\bar{h}_{v}i\not{D} _{\bot }H_{v}+\bar{H}_{v }i\not{D}_{\bot
}h_{v}  \label{3h2}
\end{equation}
Thus equation of motion for $\bar{H}_{v }$ becomes

\begin{equation}
H_{v }(x)=\frac{1}{2m_{Q}+i v \cdot D}i\not{D}_{\bot }h_{v }
\end{equation}
This allows us, on a classical level, to eliminate out the heavy degree of
freedom $H_{v}$ from the lagrangian:
\begin{eqnarray}
\mathcal{L}_{eff} &=&\bar{h}_{v }i(v \cdot D)h_{v }+\bar{h} _{v }i\not{D}%
_{\bot }\frac{1}{2m_{Q}+i v\cdot D}i\not{D}_{\bot }h_{v }  \notag \\
&=&\bar{h}_{v }i(v \cdot D)h_{v }+\frac{1}{2m_{Q}} \sum_{n=0}^{\infty }\bar{h%
}_{v }i\not{D}_{\bot }\left( -\frac{ iv \cdot D}{2m_{Q}}\right) ^{n}i\not{D}%
_{\bot }h_{v }  \label{3h3}
\end{eqnarray}
The above equation can be written as:
\begin{equation}
\mathcal{L}_{eff}=\bar{h}_{v}i(v \cdot D)h_{v }+\frac{1}{2m_{Q}}\bar{h}_{v
}\left( i\not{D}_{\bot }\right) ^{2}h_{v }+ \frac{g_{s}}{4m_{Q}}\bar{h}%
_{\upsilon }\sigma _{\mu \upsilon }G^{\mu v}h_{v }+O(1/m_{Q})^{2}
\label{3h4}
\end{equation}
where $G_{\mu\nu}$ is the gluon field strength tensor and is defined as $%
G_{\mu\nu}=[iD_{\mu},iD_{\nu}]=ig_{s}t^{a}G_{a}^{\mu\nu}$ and $%
\sigma_{\mu\nu}=\frac{1}{2}[\gamma_{\mu},\gamma_{\nu}]$.

In limit $m_{Q}\rightarrow \infty $, only term
\begin{equation}
\mathcal{L}_{\infty }=\bar{h}_{v }i(v \cdot D)h_{v}  \label{3h5}
\end{equation}
survives. There appears neither Dirac matrices nor quark masses in this
equation. For $m_{Q}\rightarrow \infty $, the interaction of heavy quarks
and gluons become independent of the spin of the quark. Furthermore, when
extending the theory to more than one heavy quark moving at the same
velocity, the lagrangian $\mathcal{L}_{\infty }\ $is symmetric under
rotations in the flavor space. This is the heavy quark flavor symmetry. The
spin-flavor symmetry leads to many interesting relations between the
properties, especially the spectroscopy, of hadrons containing a heavy
quark. In the following sections we will use the HQET to evaluate the
hadronic form factors which appear in the transition matrix.

\subsection{The Physics of Beauty(b)-Hadrons}

The hadrons which contain one beauty(bottom) b-quarks as an ingredient with
lighter quarks like, $u,\ d,\ s,$ or $c$, are called b-hadrons. As b-quark
is the heaviest quark which can be hadronized so the b-hadron can give us
rich phenomenology to understand the nature of fundamental interactions by
studying these hadrons.

The B meson is the hydrogen atom of quantum chromodynamics (QCD), the
simplest non-trivial hadron. In the leading approximation, the b-quark in it
just sits at rest at the origin and creates a chromoelectric field. Light
constituents (gluons, light quarks, and antiquarks) move in this external
field. Their motion is relativistic; the number of gluons and light
quark--antiquark pairs in this light cloud is undetermined and varying.
Similarly, the $\Lambda_{b}$ baryon is the simplest b-baryon, its quark
contents are $bud$. Both, B-meson and $\Lambda_{b}$-baryon, have a light
cloud with a variable number of relativistic particles. The size of this
cloud is the confinement radius $1/\Lambda_{QCD}$; its properties are
determined by large-distance nonperturbative QCD.

In this work, we will consider the analysis of simplest b-baryon $i.e.$ $%
\Lambda_{b}$ decay. More specifically we will consider the hadronic decay of
the type $\Lambda_{b}\to \Lambda V$, where $V$ is a vector meson with $%
j^{P}=1^{-}$, as $J/\psi$, $\rho$ or $\omega$. The effective Hamiltonian for
the decay can be given as:
\begin{equation}
\mathcal{H}^{eff}=\frac{G_{F}}{\sqrt{2}}V_{qb}V_{qs}^{\ast
}\sum_{i=1}^{10}C_{i}\left( m_{b}\right) O_{i}\left( m_{b}\right)
\end{equation}
where $C_{i}\left( m_{b}\right) $ are the Wilson Coefficients and the
operators, $O_{i}\left( m_{b}\right) $ can be understood as local operators
which govern the weak interaction of quarks in the given decay.

By using the Factorization assumption we can get the helicity amplitude for
the decay $\Lambda _{b}\rightarrow \Lambda V$ $(1^{-})$ as
\begin{equation}
A_{\left( \lambda ,\lambda ^{\prime }\right) }=\frac{G_{F}}{\sqrt{2}}
f_{V}E_{V}\langle \Lambda (p^{\prime},s^{\prime})\left\vert \bar{s}\gamma
_{\mu }\left( 1-\gamma _{5}\right) b\right\vert \Lambda _{b}(p,s)\rangle
_{\left( \lambda ,\lambda ^{\prime }\right) }\left\{
V_{CKM}^{T}C_{i}^{T}-V_{CKM}^{P}C_{i}^{P}\right\}
\end{equation}
where $f_{V}$ and $E_{V}$ are the decay constant and energy of Vector meson.
$V_{CKM}^{T,P}=V_{qb}V_{qs}^{\ast }$ are the CKM matrix elements for the
tree and penguin diagrams while $C_{i}^{T,P}$ are Wilson Coefficients. The
baryonic matrix element $\mathcal{M}_{\left( \lambda ,\lambda ^{\prime
}\right) }^{\Lambda _{b}}\equiv \langle \Lambda
(p^{\prime},s^{\prime})\left\vert \bar{s}\gamma _{\mu }\left( 1-\gamma
_{5}\right) b\right\vert \Lambda _{b}(p,s)\rangle _{\left( \lambda ,\lambda
^{\prime }\right) }$ is calculated by using the Heavy Quark Effective
Theory(HQET), in the preceding sections.

\subsection{Operator Product Expansion}

The Operator Product Expansion (OPE) $\cite{OPE}$ is used to separate the
calculation of a baryonic decay amplitude, into two distinct physical
regimes, as discussed above. One is called hard or short-distance physics,
represented by \emph{Wilson Coefficients} and the other is called soft or
long-distance physics. This part is described by $O_{i}(\mu)$, and is
derived by using a nonperturbative approach. The operators, $O_{i}$'s,
entering from the Operator Product Expansion (OPE) to reproduce the weak
interaction of quarks, can be understood as local operators which govern a
given decay. They can be written, in a generic form, as,
\begin{equation}
O_{i}=\left( \bar{q}_{\alpha }\Gamma _{i1}q_{\beta }\right) \left( \bar{q}%
_{\mu }\Gamma _{i2}q_{\nu }\right)
\end{equation}
where $\Gamma _{ij}$ denotes the gamma matrices. They should respect the
Dirac structure, the color structure and the type of quark relevant for the
decay being studied. Two kinds of topology contributing to the decay can be
defined: there is the tree diagram of which the operators are $O_{1}$, $%
O_{2} $ and the penguin diagram expressed by the operators $O_{3}$ to $O_{10}
$. The operators related to these diagrams mentioned previously are the
following,
\begin{eqnarray*}
O_{1}&=&\bar{q}_{\alpha}\gamma_{\mu}(1-\gamma_{5})u_{\beta}\bar{u}%
_{\beta}\gamma^{\mu}(1-\gamma_{5})b_{\alpha},\ \ \ \ \ \ \ \ \ \ \ \ O_{2}=%
\bar{q}\gamma_{\mu}(1-\gamma_{5})u\bar{u}\gamma^{\mu}(1-\gamma_{5})b, \\
O_{3}&=&\bar{q}\gamma_{\mu}(1-\gamma_{5})b\sum_{q^{\prime}}\bar{q}%
^{\prime}\gamma^{\mu}(1-\gamma_{5})q^{\prime},\ \ \ \ \ \ \ \ \ \ \ \ O_{4}=%
\bar{q}_{\alpha}\gamma_{\mu}(1-\gamma_{5})b_{\beta}\sum_{q^{\prime}}\bar{q}%
^{\prime}_{\beta}\gamma^{\mu}(1-\gamma_{5})q^{\prime}_{\alpha}, \\
O_{5}&=&\bar{q}\gamma_{\mu}(1-\gamma_{5})b\sum_{q^{\prime}}\bar{q}%
^{\prime}\gamma^{\mu}(1+\gamma_{5})q^{\prime},\ \ \ \ \ \ \ \ \ \ \ \ O_{6}=%
\bar{q}_{\alpha}\gamma_{\mu}(1-\gamma_{5})b_{\beta}\sum_{q^{\prime}}\bar{q}%
^{\prime}_{\beta}\gamma^{\mu}(1+\gamma_{5})q^{\prime}_{\alpha}, \\
O_{7}&=&\frac{3}{2}\bar{q}\gamma_{\mu}(1-\gamma_{5})b\sum_{q^{\prime}}e_{q^{%
\prime}}\bar{q}^{\prime}\gamma^{\mu}(1+\gamma_{5})q^{\prime},\ \ \ \ \ \ \
O_{8}=\frac{3}{2}\bar{q}_{\alpha}\gamma_{\mu}(1-\gamma_{5})b_{\beta}%
\sum_{q^{\prime}}e_{q^{\prime}}\bar{q}^{\prime}_{\beta}\gamma^{\mu}(1+%
\gamma_{5})q^{\prime}_{\alpha}, \\
O_{9}&=&\frac{3}{2}\bar{q}\gamma_{\mu}(1-\gamma_{5})b\sum_{q^{\prime}}e_{q^{%
\prime}}\bar{q}^{\prime}\gamma^{\mu}(1-\gamma_{5})q^{\prime}, \ \ \ \ \ \
O_{10}=\frac{3}{2}\bar{q}_{\alpha}\gamma_{\mu}(1-\gamma_{5})b_{\beta}%
\sum_{q^{\prime}}e_{q^{\prime}}\bar{q}^{\prime}_{\beta}\gamma^{\mu}(1-%
\gamma_{5})q^{\prime}_{\alpha},
\end{eqnarray*}
In the above expressions, $\alpha$ and $\beta$ are the color indices. $e_{q}$
denotes the quark electric charge and $q^{\prime}$, is for the quarks $u,\
d,\ c,\ s$, which may contribute in the penguin loop.

The Wilson coefficients $\cite{OPE}$, $C_{i}(\mu)$, represent the physical
contributions from scales higher than $\mu$ (of the order of $O(m_{b})$ in
b-quark decay) and since QCD has the property of asymptotic freedom, they
can be calculated in perturbation theory. we taken the Wilson coefficients
from $\cite{ajal}$ for $q^2/m_{b}^{2}=0.5$ and their values are summarized
as:
\begin{table}[ht]
\caption{Wilson coefficients for tree and penguin operators}\centering
\begin{tabular}{cccc}
\hline\hline
$C_1$ & $-0.3125$ & $C_2$ & $1.1502$ \\ \hline
$C_3$ & $2.12\times10^{-2}+i 2.174\times10^{-3}$ & $C_4$ & $%
-4.869\times10^{-2}-i 1.552\times10^{-2}$ \\ \hline
$C_5$ & $1.42\times10^{-2}+i 5.174\times10^{-3}$ & $C_6$ & $%
-5.729\times10^{-2}-i 1.552\times10^{-2}$ \\ \hline
$C_7$ & $-8.34\times10^{-5}-i 9.94\times10^{-5}$ & $C_8$ & $%
3.84\times10^{-4} $ \\ \hline
$C_9$ & $-1.02\times10^{-2}-i 9.94\times10^{-5}$ & $C_{10}$ & $%
1.96\times10^{-3}$ \\ \hline\hline
\end{tabular}%
\end{table}

Finally, in the following one lists the tree and penguin amplitudes which
appear in the given transition: \newline
for the decay$\Lambda_{b}\to J/\psi$,
\begin{equation}
A_{J/\psi}^{T}(a_{1},a_{2})=a_{1}
\end{equation}
\begin{equation}
A_{J/\psi}^{P}(a_{3},...,a_{10})=a_{3}+a_{5}+a_{7}+a_{9}
\end{equation}
\newline
for the decay$\Lambda_{b}\to \rho^{0}$,
\begin{equation}
A_{\rho}^{T}(a_{1},a_{2})=\frac{a_{1}}{\sqrt{2}}
\end{equation}
\begin{equation}
A_{\rho}^{P}(a_{3},...,a_{10})=\frac{3}{2\sqrt{2}}%
(4(a_{3}+a_{5})+a_{7}+a_{9})
\end{equation}
\newline
for the decay$\Lambda_{b}\to \omega$,
\begin{equation}
A_{\omega}^{T}(a_{1},a_{2})=\frac{a_{1}}{2\sqrt{2}}
\end{equation}
\begin{equation}
A_{\omega}^{P}(a_{3},...,a_{10})=\frac{3}{2\sqrt{2}}(a_{7}+a_{9})
\end{equation}
where $a_{i}=C_{i}+C_{j}/N_{c}$ with $i,j=1,2,...,10$ and $N_{c}$ is the
number of colors.

\subsection{Evolution of Baryonic Form Factors in HQET}

In this section, the Heavy Quark Effective Theory (HQET) formalism is used
to evaluate the hadronic form factors involved in $\Lambda_{b}$-decay. Weak
transitions including heavy quarks can be safely described when the mass of
a heavy quark is large enough compared to the QCD scale, $\Lambda_{QCD}$.
Properties such as flavor and spin symmetries can be exploited in such way
that corrections of the order of $1/m_{Q}$ are systematically calculated
within an effective field theory.

\subsubsection{Transition Form Factors}

The decay, $\Lambda_{b}\rightarrow\Lambda V$, involves the hadronic
transition matrix $\langle \Lambda \left\vert \bar{s}\gamma _{\mu }\left(
1-\gamma _{5}\right) b\right\vert \Lambda _{b}\rangle$. Based on Lorentz
decomposition, the hadronic matrix element can be written as,
\begin{equation}
\langle \Lambda(p^{\prime},s^{\prime}) \left\vert \bar{s}\gamma _{\mu
}\left( 1-\gamma _{5}\right) b\right\vert \Lambda _{b}(p,s)\rangle= \bar{u}%
_{\Lambda}(p^{\prime},s^{\prime})\left\{%
\begin{array}{c}
\left( f_{1}(q^{2})\gamma_{\mu}+i
f_{2}(q^{2})\sigma_{\mu\nu}q^{\nu}+f_{3}(q^{2})q_{\mu} \right) \\
-\left( g_{1}(q^{2})\gamma_{\mu}+i
g_{2}(q^{2})\sigma_{\mu\nu}q^{\nu}+g_{3}(q^{2})q_{\mu} \right)\gamma_{5}%
\end{array}%
\right\}u_{\Lambda_{b}}(p,s)  \label{3f1}
\end{equation}
where $\bar{u}_{\Lambda}(p^{\prime},s^{\prime})$ and $u_{\Lambda_{b}}(p,s)$
are the spinners of $\Lambda$ and $\Lambda_{b}$ respectively, while $%
p^{\prime},s^{\prime}$ and $p,s$ are their momentum and spin. The square of
momentum transfer in the hadronic transition is given by
\begin{equation*}
q^{2}=(p-q^{\prime})^{2}
\end{equation*}
Here $f_{i}(q^{2})$ and $g_{i}(q^{2})$ are the form factors corresponding to
the vector and axial vector parts of the transition matrix, respectively.

Another way of parameterizing the electroweak amplitude in decays of baryons
is the following:
\begin{equation}
\langle \Lambda(p^{\prime},s^{\prime}) \left\vert \bar{s}\gamma _{\mu
}\left( 1-\gamma _{5}\right) b\right\vert \Lambda _{b}(p,s)\rangle= \bar{u}%
_{\Lambda}(p^{\prime},s^{\prime})\left\{
\begin{array}{c}
\left(F_{1}(q^{2})\gamma_{\mu}+ F_{2}(q^{2})v_{\mu\Lambda_{b}}+F_{3}(q^{2})%
\frac{P^{\prime}_{\mu}}{m^{\prime}} \right) \\
-\left(G_{1}(q^{2})\gamma_{\mu}+G_{2}(q^{2})v_{\mu\Lambda_{b}}+G_{3}(q^{2})%
\frac{P^{\prime}_{\mu}}{m^{\prime}} \right)\gamma_{5}%
\end{array}%
\right\}u_{\Lambda_{b}}(p,s)  \label{3f2}
\end{equation}
By comparing the two sets of form factors given in Eqs. $\eqref{3f1}$ and $%
\eqref{3f2}$, we gets the following relations between the $f_{i}(q^{2})$'s$%
(g_{i}(q^{2})^{\prime }s)$ and $F_{i}(q^{2})$'s$(f_{i}(q^{2})^{\prime }s)$:
\begin{eqnarray}
f_{1}(q^{2})&=&F_{1}(q^{2})+(m+m^{\prime})\left[ \frac{F_{2}(q^{2})}{2m}+%
\frac{F_{3}(q^{2})}{2m^{\prime}}\right], \\
f_{2}(q^{2})&=&\frac{F_{2}(q^{2})}{2m}+\frac{F_{3}(q^{2})}{2m^{\prime}}, \\
f_{3}(q^{2})&=&\frac{F_{2}(q^{2})}{2m}-\frac{F_{3}(q^{2})}{2m^{\prime}}
\label{3f3}
\end{eqnarray}
and
\begin{eqnarray}
g_{1}(q^{2})&=&G_{1}(q^{2})-(m-m^{\prime})\left[ \frac{G_{2}(q^{2})}{2m}+%
\frac{G_{3}(q^{2})}{2m^{\prime}}\right], \\
g_{2}(q^{2})&=&\frac{G_{2}(q^{2})}{2m}+\frac{G_{3}(q^{2})}{2m^{\prime}}, \\
g_{3}(q^{2})&=&\frac{G_{2}(q^{2})}{2m}-\frac{G_{3}(q^{2})}{2m^{\prime}}
\label{3f4}
\end{eqnarray}
In case of working in the HQET formalism, the matrix element of the weak
transition, $\Lambda_{b}\rightarrow\Lambda$, takes the following form,
\begin{equation}
\langle \Lambda(p^{\prime},s^{\prime}) \left\vert \bar{s}\gamma _{\mu
}\left( 1-\gamma _{5}\right) b\right\vert \Lambda _{b}(p,s)\rangle= \bar{u}%
_{\Lambda}(p^{\prime},s^{\prime})\left[ \theta_{1}(q^{2})+ \theta_{2}(q^{2})%
\not{v}_{\Lambda_{b}}\right]u_{\Lambda_{b}}(p,s)  \label{3f5}
\end{equation}
In Eq. $\eqref{3f5}$, $v_{\Lambda_{b}}$, defines the velocity of the baryon $%
\Lambda_{b}$. Writing the momentum, $p$ , of the heavy baryon, $\Lambda_{b}$
as,
\begin{equation*}
p=m_{b}v_{\Lambda_{b}}+k,
\end{equation*}
where $k$ is the residual momentum, the velocity of heavy quark is almost
that of the heavy baryon. Since $m_{b}\gg\Lambda_{QCD}$, the parametrization
of the hadronic matrix element in term of velocity, $v_{\Lambda_{b}}$, gives
us a reasonable picture where we can consider only corrections of $1/m_{b}$
expansion.

Since we know that in heavy hadrons the spectator quark retains its original
momentum and spin state before final hadronization, the energy carried by
the spectator quark is equal to that of the spectator in the rest frame of
the final state particle and the relevant $b$-quark space momenta are much
smaller than the $b$ quark mass: indeed, it is assumed to be of the order of
the confinement scale, $\Lambda_{QCD}$. This approach firstly used in the
meson case by Stech but can be generalized to a heavy baryon considered as a
bound state of a b quark and a scalar diquark as considered in $\cite{ghl}$.
Thus in the baryon case hadronic matrix can be written in terms of
components of Dirac Spinors as, $\bar{u}_{s}(p^{\prime},m_{s})\gamma _{\mu
}\left( 1-\gamma _{5}\right)u_{b}(p=0,m_{b})$ leads to the following
expressions for the form factors, $\theta_{1}$ and $\theta_{2}$, when the $%
m_{b}\rightarrow \infty$:
\begin{eqnarray}
\theta_{1}&=&\left( E_{\Lambda}+m^{\prime}+m_{s}\right)\frac{1}{%
(E_{\Lambda}+m_{s})}\sqrt{\frac{(E_{\Lambda}+m_{s})m^{\prime}}{%
(E_{\Lambda}+m^{\prime})m_{s}}}  \label{3f6} \\
\theta_{2}&=&\left( m_{s}-m^{\prime}\right)\frac{1}{2(E_{\Lambda}+m_{s})}%
\sqrt{\frac{(E_{\Lambda}+m_{s})m^{\prime}}{(E_{\Lambda}+m^{\prime})m_{s}}}
\label{3f7}
\end{eqnarray}
where $E_{\Lambda}$, is the energy of $\Lambda$ in the rest frame of $%
\Lambda_{b}$, and is given by:
\begin{equation}
E_{\Lambda}=\frac{m^{2}+m^{\prime2}-q^{2}}{2m}  \label{3f8}
\end{equation}
Here $m$ and $m^{\prime}$ are the masses of $\Lambda_{b}$ and $\Lambda$,
respectively, with $q^{2}$ as described above. It is convenient to define
the invariant velocity transfer, $\omega(q^{2})$, as
\begin{equation}
\omega(q^{2})=v\cdot v^{\prime}= \frac{m^{2}+m^{\prime2}-q^{2}}{2m m^{\prime}%
}  \label{3f9}
\end{equation}
where $v$ and $v^{\prime}$ are the four velocities of $\Lambda_{b}$ and $%
\Lambda$. The minimum and maximum values of $\omega(q^{2})$ are obtained
corresponding to $q^{2}=(m-m^{\prime})^{2}$ and $q^{2}=0$ as
\begin{equation*}
\omega_{min}(q^{2})=1,\ \ \ \ \ \omega_{max}(q^{2})=\frac{m^{2}+m^{\prime2}}{%
2m m^{\prime}}
\end{equation*}
The zeroth order form factors $F^{0}_{i}$'s and $G^{0}_{i}$'s in terms of $%
\theta_{1}$ and $\theta_{2}$ are given as:
\begin{equation}
F_{1}^{0}=\theta_{1}-\theta_{2},\ \ \ \ \ \ F_{2}^{0}= 2 \theta_{2},\ \ \ \
\ \ F_{3}^{0}=0,  \label{3fz1}
\end{equation}
and
\begin{equation}
G_{1}^{0}=\theta_{1}+\theta_{2},\ \ \ \ \ \ G_{2}^{0}= 2 \theta_{2},\ \ \ \
\ \ G_{3}^{0}=0,  \label{3fz2}
\end{equation}
These zeroth order form factors lead to the following relations,
\begin{equation}
G_{1}^{0}=F_{1}^{0}+F_{2}^{0}; \ \ \ \ \ G_{2}^{0}=F_{2}^{0}; \ \ \ \ \
G_{3}^{0}=F_{3}^{0}=0  \label{3f10}
\end{equation}
or equivalently,
\begin{equation}
g_{1}=f_{1}; \ \ \ \ \ g_{2}=f_{2}; \ \ \ \ \ g_{3}=f_{3}=-f_{2}
\label{3f11}
\end{equation}
The radiative corrections will not be taken into account since they are not
relevant in our analysis whereas the corrections proportional to $%
\Lambda_{QCD}/m_{b}$ will be systematically calculated. These latter
nonperturbative corrections are computed in the next section. In the
following, all the form factors will be defined as a function of the
invariant velocity transfer, $\omega(q^{2})$, instead of the momentum
transfer, $q^{2}$.

\subsubsection{$1/m_{b}$ Corrections to the Form Factors}

Since the effective lagrangian in HQET is given by
\begin{equation}
\mathcal{L}_{\infty }=\bar{h}_{v }i(v \cdot D)h_{v}
\end{equation}
where $h_{v}$ is the quark field as defined in the previous section, it
corresponds to $b$-quark in our case. Including the corrections of $1/m_{b}$
the effective Lagrangian has the form,
\begin{equation}
\mathcal{L}_{eff}=\bar{h}_{v}i(v \cdot D)h_{v }+\frac{1}{2m_{Q}}\bar{h}_{v
}\left( i\not{D}_{\bot }\right) ^{2}h_{v }+ \frac{g_{s}}{4m_{Q}}\bar{h}%
_{\upsilon }\sigma _{\mu \upsilon }G^{\mu v}h_{v }+O(1/m_{Q})^{2}
\end{equation}

In case of heavy to light quarks mass transition, the weak current will have
the following general structure, up to the $1/m_{b}$ corrections;
\begin{equation}
\bar{q}\Gamma\psi_{(b)}\to \bar{q}\Gamma h_{v}+\frac{1}{m_{b}}\bar{q}i\not{D}%
h_{v}+O(1/m_{Q})^{2}  \label{3f12}
\end{equation}
where $\Gamma$ can have values $\gamma_{\mu}$ or $\gamma_{\mu}\gamma_{5}$.

By including the covariant derivative, $D$, as well as the corrections at
the order of $1/m_{b}$ to the effective Lagrangian, it leads, respectively,
to the local correction given by,
\begin{equation}
\delta\mathcal{L}^{lo,1}=\frac{1}{2m_{b}}\bar{q}\Gamma i\not{D}h_{v}
\label{3f13}
\end{equation}
and the non-local corrections given by,
\begin{eqnarray}
\delta\mathcal{L}^{nlo,2}&=&\frac{1}{2m_{b}}\bar{h_{v}}(iv \cdot D)^{2}h_{v},
\\
\delta\mathcal{L}^{nlo,3}&=&\frac{1}{2m_{b}}\bar{h_{v}}(i D)^{2}h_{v}, \\
\delta\mathcal{L}^{nlo,4}&=&\frac{1}{2m_{b}}\bar{h_{v}}\sigma_{\mu\nu}G^{\mu%
\nu} h_{v}
\end{eqnarray}
where $q$ stands for the light quarks $u$, $d$ or $s$.

Let us start with the local term correction, $\delta\mathcal{L}^{lo,1}$, to
the effective lagrangian. The matrix element usually takes the form,
\begin{equation}
\langle \Lambda(p^{\prime},s^{\prime}) \left\vert \bar{q}\Gamma iD
h_{v}\right\vert \Lambda _{b}(p,s)\rangle= \bar{u}_{\Lambda}(p^{\prime},s^{%
\prime})\phi^{\mu}(\omega)\Gamma \gamma_{\mu}u_{\Lambda_{b}}(p,s)
\label{3f14}
\end{equation}
where the form of the R.H.S of the above equation fallow from the spin
symmetry. The most general form of $\phi^{\mu}$ is,
\begin{equation}
\phi^{\mu}=(\phi_{11}v^{\mu}+\phi_{12}v^{\prime\mu}+\phi_{13}\gamma^{\mu})+%
\not{v}(\phi_{21}v^{\mu}+\phi_{22}v^{\prime\mu}+\phi_{23}\gamma^{\mu})
\label{3f15}
\end{equation}
On the other hand, the equation of motion for heavy quark is,
\begin{equation}
v\cdot D h_{v}=0  \label{3f16}
\end{equation}
When it is applied on eq.$\eqref{3f14}$, we get,
\begin{equation}
v \cdot \langle \Lambda(p^{\prime},s^{\prime}) \left\vert \bar{q}\Gamma iD
h_{v}\right\vert \Lambda _{b}(p,s)\rangle= 0  \label{3f17}
\end{equation}
and it leads therefore to the following constraints, for $\Gamma=1$ and $%
\Gamma=\gamma_{5}$, respectively,
\begin{eqnarray}
\bar{u}_{\Lambda}(p^{\prime},s^{\prime})[v\cdot\phi(\omega)]u_{%
\Lambda_{b}}(p,s)=0 \\
\bar{u}_{\Lambda}(p^{\prime},s^{\prime})[v\cdot\phi(\omega)\gamma_{5}]u_{%
\Lambda_{b}}(p,s)=0
\end{eqnarray}
Thus, two relations between the $\phi_{ij}$'s can be obtained from the above
constraints as,
\begin{eqnarray}
\phi_{11}+\omega\phi_{12}=-\phi_{23} \\
\phi_{21}+\omega\phi_{22}=-\phi_{13}
\end{eqnarray}
On the other hand, the momentum conservation also implies that,
\begin{eqnarray}
\langle \Lambda(p^{\prime},s^{\prime}) \left\vert i \partial_{\mu}( \bar{q}%
\Gamma i D h_{v})\right\vert \Lambda _{b}(p,s)\rangle &=& \langle
\Lambda(p^{\prime},s^{\prime}) \left\vert iD_{\mu}\bar{q}\Gamma
h_{v}\right\vert \Lambda _{b}(p,s)\rangle + \langle
\Lambda(p^{\prime},s^{\prime}) \left\vert \bar{q}\Gamma i D_{\mu}
h_{v}\right\vert \Lambda _{b}(p,s)\rangle  \notag \\
&=&\left\{(m-m^{\prime})v_{\mu}-m^{\prime}v^{\prime}_{\mu} \right\}\langle
\Lambda(p^{\prime},s^{\prime}) \left\vert \bar{q}\Gamma iD h_{v}\right\vert
\Lambda _{b}(p,s)\rangle
\end{eqnarray}
where the equation of motion for light quark, $(i\not D-m_{q})q=0$ has been
used. Changing $\Gamma$ by $\gamma_{\mu}\Gamma$, so that $\Gamma$ is limited
to $1$ and $\gamma_{5}$ for vector and axial vector currents. We get,
\begin{equation}
\langle \Lambda(p^{\prime},s^{\prime}) \left\vert \bar{q}\gamma^{\mu}\Gamma
i D_{\mu} h_{v}\right\vert \Lambda
_{b}(p,s)\rangle=\left\{(m-m^{\prime})v_{\mu}-(m^{\prime}-m_{q})v^{\prime}_{%
\mu} \right\}\langle \Lambda(p^{\prime},s^{\prime}) \left\vert \bar{q}%
\gamma^{\mu}\Gamma iD h_{v}\right\vert \Lambda _{b}(p,s)\rangle
\end{equation}
For $\Gamma=1$ and $\Gamma=\gamma_{5}$, from the above equation we get,
respectively,
\begin{equation}
\left[\frac{\omega-1}{\omega}\right](\phi_{11}-\phi_{12})-\left[\frac{%
2\omega+1}{\omega}\right]\phi_{13}+\left[\frac{4\omega-1}{\omega}\right]%
\phi_{23}=(m-m_{b})(F_{1}^{0}+F_{2}^{0})-(m^{\prime}-m_{q})(F_{1}^{0}+\omega
F_{2}^{0}),  \label{3f18}
\end{equation}
and
\begin{equation}
\left[\frac{\omega+1}{\omega}\right](\phi_{11}-\phi_{12})-\left[\frac{%
2\omega-1}{\omega}\right]\phi_{13}+\left[\frac{4\omega-1}{\omega}\right]%
\phi_{23}=(m-m_{b})(G_{1}^{0}-G_{2}^{0})-(m^{\prime}-m_{q})(G_{1}^{0}+\omega
G_{2}^{0}),  \label{3f19}
\end{equation}
In the above equations, $F_{i}^{0}$'s and $G_{i}^{0}$'s are the zeroth order
form factors as given by eqs.$(\ref{3fz1},\ref{3fz2})$.

We can get the expressions for $\phi_{ij}$'s from the above eqs. $%
\eqref{3f18}$ and $\eqref{3f19}$ in terms of zeroth order form factors as:
\begin{eqnarray}  \label{3f22}
\phi_{11}(\omega)&=&\frac{\omega(\omega+1)}{2(\omega^{2}-1)}\left[
(m-m_{b})(F_{1}^{0}+F_{2}^{0})-(m^{\prime}-m_{q})(F_{1}^{0}+\omega F_{2}^{0})%
\right]  \notag \\
&&+\frac{\omega(\omega-1)}{2(\omega^{2}-1)}\left[
(m-m_{b})(G_{1}^{0}-G_{2}^{0})+(m^{\prime}-m_{q})(G_{1}^{0}+\omega G_{2}^{0})%
\right]-\frac{7\omega-1}{\omega^{2}-1}\phi_{123}(\omega)  \notag \\
\\
\phi_{12}(\omega)&=&\frac{\omega-1}{2(\omega^{2}-1)}\left[
-(m-m_{b})(F_{1}^{0}+F_{2}^{0})+(m^{\prime}-m_{q})(F_{1}^{0}+\omega
F_{2}^{0})\right]  \notag \\
&&+\frac{\omega-1}{2(\omega^{2}-1)}\left[
-(m-m_{b})(G_{1}^{0}-G_{2}^{0})-(m^{\prime}-m_{q})(G_{1}^{0}+\omega
G_{2}^{0})\right]-\frac{\omega-7}{\omega^{2}-1}\phi_{123}(\omega)  \notag \\
\\
\phi_{21}(\omega)&=&\frac{\omega(\omega+1)}{2(\omega^{2}-1)}\left[
-(m-m_{b})(F_{1}^{0}+F_{2}^{0})+(m^{\prime}-m_{q})(F_{1}^{0}+\omega
F_{2}^{0})\right]  \notag \\
&&+\frac{\omega(\omega-1)}{2(\omega^{2}-1)}\left[
(m-m_{b})(G_{1}^{0}-G_{2}^{0})+(m^{\prime}-m_{q})(G_{1}^{0}+\omega G_{2}^{0})%
\right]+\frac{6\omega^{2}-\omega+1}{\omega^{2}-1}\phi_{123}(\omega)  \notag
\\
\end{eqnarray}
and
\begin{eqnarray}  \label{3f23}
\phi_{22}(\omega)&=&-\frac{\omega+1}{2(\omega^{2}-1)}\left[
-(m-m_{b})(F_{1}^{0}+F_{2}^{0})+(m^{\prime}-m_{q})(F_{1}^{0}+\omega
F_{2}^{0})\right]  \notag \\
&&+\frac{\omega-1}{2(\omega^{2}-1)}\left[
(m-m_{b})(G_{1}^{0}-G_{2}^{0})+(m^{\prime}-m_{q})(G_{1}^{0}+\omega G_{2}^{0})%
\right]+\frac{1-7\omega}{\omega^{2}-1}\phi_{123}(\omega)  \notag \\
\end{eqnarray}
In the above equations we have used the assumption of $\phi_{13}(\omega)%
\approx\phi_{23}(\omega)\equiv\phi_{123}(\omega)$ since they are equal at
zeroth order and are negligible at the first order corrections, which is of
the order of $1/m_{b}$, as discussed below.

The basic assumption involved in such an analysis is the following; in HQET
on the scale of the heavy quark mass the light degrees of freedom have small
momentum spread about their central equal velocity value. For strange baryon
or meson this is not true. However, it is possible that the smearing of the
momentum of the light degrees averages out effectively. In the limit of
equal hadron masses we would then have the normalization condition at $%
\omega =1$,
\begin{equation}
F_{1}+F_{2}+F_{3}=1
\end{equation}
which implies,
\begin{equation}
F_{1}^{0}+F_{2}^{0}=1  \label{3f24a}
\end{equation}
We therefore get the condition on corrections to the form factors as
\begin{equation}
\delta F_{1}+\delta F_{2}+\delta F_{3}=0  \label{3f24}
\end{equation}
in the limit of equal hadron masses.

In this work we do not assume the validity of an $1/m_{s}$ expansion but we
make the assumption that the eq.$\eqref{3f24}$ is valid upto to the order we
are working in even for unequal hadron masses or at most the R.H.S of eq.$%
\eqref{3f24}\sim \epsilon/2m_{b}$ for unequal hadron masses. This is indeed
the case in heavy to heavy transitions where for example both eqns.$%
\eqref{3f24}$ and $\eqref{3f24a}$ are true for $\Lambda_{b}\to \Lambda_{c}$
upto $1/m_{Q}^{2}$ for unequal hadron masses and it is a consequence of
Luke's theorem. So in our case we have unequal masses of hadron so we have,
\begin{equation}
\delta F_{1}^{lo,1}+\delta F_{2}^{lo,1}+\delta F_{3}^{lo,1}=\frac{\epsilon}{%
2m_{b}}  \label{3f25}
\end{equation}
This allows us to derive the expression of $\phi_{123}(\omega)$ as,
\begin{eqnarray}
\phi_{123}(\omega)&=&\frac{\omega+1}{16(\omega-1)}\left[ \epsilon
+(m-m_{b})(F_{1}^{0}+F_{2}^{0})-(m^{\prime}-m_{q})(F_{1}^{0}+\omega
F_{2}^{0})\right]  \notag \\
&&+\frac{1}{8}\left[ -(m-m_{b})(G_{1}^{0}-G_{2}^{0})-(m^{%
\prime}-m_{q})(G_{1}^{0}+\omega G_{2}^{0})+\frac{\epsilon(\omega+1)}{%
2(\omega-1)}\right]  \label{3f26}
\end{eqnarray}

It is now obvious to calculate the local corrections to the form factors as;
\begin{eqnarray}
\delta F_{1}^{lo,1}(\omega)&=&-\frac{1}{2m_{b}}\left[\phi_{11}(\omega)+(2%
\omega+1)\phi_{12}(\omega)-\phi_{21}(\omega)+\phi_{22}(\omega)\right] \\
\delta F_{2}^{lo,1}(\omega)&=&\frac{1}{m_{b}}\left[2\phi_{11}(\omega)+2%
\omega\phi_{12}(\omega)+\phi_{21}(\omega)+\phi_{22}(\omega)\right] \\
\delta F_{3}^{lo,1}(\omega)&=&\frac{1}{m_{b}}\left[\phi_{11}(\omega)+%
\phi_{21}(\omega)\right]
\end{eqnarray}
and
\begin{eqnarray}
\delta G_{1}^{lo,1}(\omega)&=&\frac{1}{2m_{b}}\left[\phi_{11}(\omega)+(2%
\omega-1)\phi_{12}(\omega)-\phi_{21}(\omega)+\phi_{22}(\omega)\right] \\
\delta G_{2}^{lo,1}(\omega)&=&\frac{1}{m_{b}}\left[2\phi_{11}(\omega)+2%
\omega\phi_{12}(\omega)-\phi_{21}(\omega)+\phi_{22}(\omega)\right] \\
\delta G_{3}^{lo,1}(\omega)&=&\frac{1}{m_{b}}\left[\phi_{12}(\omega)-%
\phi_{22}(\omega)\right]
\end{eqnarray}
We can safely neglect the non-local corrections to the form factors because $%
\delta\mathcal{L}^{nlo,2}=\frac{1}{2m_{b}}\bar{h_{v}}(iv \cdot D)^{2}h_{v},
\delta\mathcal{L}^{nlo,3}=\frac{1}{2m_{b}}\bar{h_{v}}(i D)^{2}h_{v},\text{
and } \delta\mathcal{L}^{nlo,4}=\frac{1}{2m_{b}}\bar{h_{v}}%
\sigma_{\mu\nu}G^{\mu\nu} h_{v}$ will only appear at the order of $%
1/m_{b}^{2}$ and such these have a negligible contributions.

Thus the full form factors after incorporating the $1/m_{b}$ corrections,
are
\begin{eqnarray}
F_{i}(\omega)&=&F_{i}^{0}+\delta F_{i} \\
G_{i}(\omega)&=&G_{i}^{0}+\delta G_{i}
\end{eqnarray}
Explicitly, we can write the expressions of the form factors as
\begin{eqnarray}
F_{1}(\omega)&=&F_{1}^{0}(\omega)-\frac{1}{2m_{b}}\left[\phi_{11}(\omega)+(2%
\omega+1)\phi_{12}(\omega)-\phi_{21}(\omega)+\phi_{22}(\omega)\right] \\
F_{2}(\omega)&=&F_{2}^{0}(\omega)+\frac{1}{m_{b}}\left[2\phi_{11}(\omega)+2%
\omega\phi_{12}(\omega)+\phi_{21}(\omega)+\phi_{22}(\omega)\right] \\
F_{3}(\omega)&=&F_{3}^{0}(\omega)+\frac{1}{m_{b}}\left[\phi_{11}(\omega)+%
\phi_{21}(\omega)\right]
\end{eqnarray}
and,
\begin{eqnarray}
G_{1}(\omega)&=&G_{1}^{0}+\frac{1}{2m_{b}}\left[\phi_{11}(\omega)+(2%
\omega-1)\phi_{12}(\omega)-\phi_{21}(\omega)+\phi_{22}(\omega)\right] \\
G_{2}(\omega)&=&G_{2}^{0}+\frac{1}{m_{b}}\left[2\phi_{11}(\omega)+2\omega%
\phi_{12}(\omega)-\phi_{21}(\omega)+\phi_{22}(\omega)\right] \\
G_{3}(\omega)&=&G_{3}^{0}+\frac{1}{m_{b}}\left[\phi_{12}(\omega)-\phi_{22}(%
\omega)\right]
\end{eqnarray}
The computational work and the evolution of the form factors verses the
invariant velocity transfer, $\omega$, are done in last chapter.

\section{Analysis of Discrete Symmetries in $\Lambda_{b}\rightarrow\Lambda
V(1^{-})$}

Looking for discrete symmetry violation effects, in b-baryon decays, can
provide us a new field of research. Especially time-reversal (TR) violation
effects can be of great interest. Firstly, TR can be seen as a complementary
test of CP violation by assuming the correctness of the CPT theorem.
Secondly, this can also be a path to follow in order to search for processes
beyond the Standard Model. So that $\Lambda_{b}$-decay seems to be one of
the most promising channel to reveal TR violation and CP violation signal.

A general formulation based on the M. Jacob- G.C. Wick-J.D. Jackson (JWJ)
helicity formalism has been set for studying the decay process $%
\Lambda_{b}\rightarrow\Lambda V(1^{-})$. Emphasis is put on the importance
of the initial $\Lambda_{b}$ polarization as well as the correlations among
the angular distributions of the final decay products. On the dynamical
side, the Hadronic Matrix Elements (HME) appearing in the decay amplitude
were computed, at the tree level approximation, in the framework of the
factorization ansatze for two-body non-leptonic weak decay of heavy quark.

\subsection{Kinematical properties of $\Lambda _{b}\rightarrow \Lambda V$ decays%
}

The hyperons produced in proton-proton collisions as well as in other hadron
collisions are usually polarized in the transverse direction. The average
value of the hyperon spin being non equal to zero and, owing to Parity
conservation in strong interaction, the spin direction is orthogonal to the
production plane defined by the incident beam momentum, $\vec{P}_{p}$, and
the hyperon momentum, $\vec{P}_{h}$. Usually, the degree of polarization
depend on the centre of mass energy and the hyperon transverse momentum. We
define $\vec{e}_{z}$ as the normal vector to the production plane:
\begin{equation*}
\vec{e}_{z}=\vec{n}=\frac{\vec{p}_{p}\times \vec{p}_{h}}{\left\vert \vec{p}
_{p}\times \vec{p}_{b}\right\vert }
\end{equation*}
Here $\vec{p}_{p}$ and $\vec{p}_{h}$ are, respectively, the proton momentum
and the hyperon momentum.

Let $(\Lambda_{b}XYZ)$ be the rest frame (See Fig 4.1) of the $\Lambda_{b}$
particle. The quantization axis $\vec{n}$ is chosen to be parallel to $\vec{e%
}_{z}$ . The other orthogonal axis $\vec{e}_{x}$ and $\vec{e}_{y}$ are
arbitrary in the production plane.
\begin{figure}[h]
\centering
\includegraphics[scale=.3]{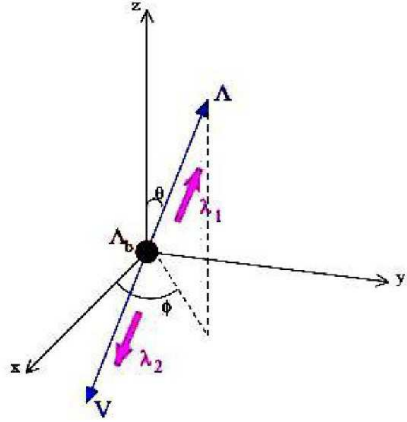} 
\caption{$\Lambda_{b}$ decay in its transversity frame}
\end{figure}

We study the kinematical properties of decays $\Lambda _{b}\rightarrow
\Lambda V$ by the Jacob-Wick-Jackson helicity formalism, since helicity
formalism has some advantages:

\begin{enumerate}
\item Helicity, $\lambda =\frac{\vec{\sigma}.\vec{p}}{\left\vert \vec{p}
\right\vert },$ depends on spin $\vec{s}$ and momentum $\vec{p}$ of the
particle and does not depend on its orbital angular momentum $\vec{\ell}$,
so that it is rotationally invariant.

\item We can work easily in the rest frame of resonances in this formalism.
\end{enumerate}

It is more convenient to define a frame of three mutually orthogonal unit
vectors
\begin{equation*}
\vec{e}_{z}=\vec{n}=\frac{\vec{p}_{p}\times \vec{p}_{b}}{\left\vert \vec{p}
_{p}\times \vec{p}_{b}\right\vert };\ \ \ \ \vec{e}_{x}=\frac{\vec{p}_{p}}{
\left\vert \vec{p}_{p}\right\vert };\ \ \ \ \vec{e}_{y}=\vec{e}_{z}\times
\vec{e}_{x}
\end{equation*}
If $\Lambda_{b}$ produced by means of strong interactions then it is
polarized along $\vec{n}$. Therefore we choose the quantization axis along $%
\vec{e}_{z}=\vec{n}.$ $\Lambda _{b}$ being transversally polarized, its
polarization value is given by $\vec{\mathcal{P}}^{\Lambda _{b}}=\langle
\vec{S}_{\Lambda b}$\textperiodcentered $\vec{e}_{z}\rangle $. Let $M_{i}$
be the $\Lambda _{b}$ spin projection along $\vec{e}_{z}$ axis.

We define the Spin Density Matrix (SDM) for $\Lambda _{b}$ as:
\begin{equation}
\rho ^{\Lambda _{b}}=\frac{1}{2}\left( 1+\vec{\mathcal{P}}^{\Lambda
_{b}}\cdot \vec{\sigma}\right)
\end{equation}
where $\vec{\sigma}=\left( \sigma _{x},\ \sigma _{y},\ \sigma _{z}\right) $
are the Pauli spin matrices.

In the rest frame of $\Lambda _{b}$ the components of its polarization
vector are
\begin{equation*}
P_{z}^{\Lambda _{b}}=\frac{1}{2}\left( \rho _{++}^{\Lambda _{b}}-\rho
_{--}^{\Lambda _{b}}\right) ,\ \ \ P_{x}^{\Lambda _{b}}=\Re \left( \rho
_{+-}^{\Lambda _{b}}\right) ,\ \ \ P_{y}^{\Lambda _{b}}=-\Im \left( \rho
_{+-}^{\Lambda _{b}}\right)
\end{equation*}

$\rho _{MM^{\prime }}^{\Lambda _{b}}$ are the matrix elements of $\rho
^{\Lambda _{b}}$; $M,M^{\prime }=\pm $ denoting the values of the third
component of the $\Lambda _{b}$ spin along the quantization axis $\vec{e}%
_{z} $. $\rho ^{\Lambda _{b}}$ verifies the normalization condition
\begin{equation*}
Tr\left( \rho ^{\Lambda _{b}}\right) =\left( \rho _{++}^{\Lambda _{b}}+\rho
_{--}^{\Lambda _{b}}\right) =1
\end{equation*}

In the framework of the JWJ helicity formalism the decay amplitude, $%
A_{0}(M_{i})$, for $\Lambda_{b}(M_{i})\rightarrow\Lambda(\lambda_{1})V(%
\lambda_{2})$ is obtained by applying the Wigner-Eckart theorem to the $%
\mathcal{S}$-matrix element:
\begin{equation}
A_{0}(M_{i}) = \langle p,\theta,\phi;\lambda_{1},\lambda_{2} |\mathcal{S}%
^{0}|1/2, M_{i}\rangle = \mathcal{A}_{(\lambda _{1,}\lambda
_{2})}D_{M_{i}M_{f}}^{1/2*}(\phi,\theta,0)
\end{equation}
where $\vec{p}=(p,\theta,\phi)$ is the momentum of the hyperon $\Lambda$ in
the $\Lambda_{b}$ frame (Fig 4.1) and the Wigner matrix is given by
\begin{equation*}
D_{M_{i}M_{f}}^{1/2*}(\phi,\theta,0)=d^{j}_{M_{i}M_{f}}(\theta)\text{exp}(
-iM_{i}\phi)
\end{equation*}
$\lambda_{1}$ and $\lambda_{2}$ are the respective helicities of $\Lambda$
and $V$ with the possible value $\lambda_{1}=\pm 1/2$ and $%
\lambda_{2}=-1,0,+1$. If $M_{i}$ is the helicity of $\Lambda_{b}$ and has
the values $\pm 1/2$ then by conservation of angular momentum we have four
possible values for the pair $\left( \lambda _{1},\lambda _{2}\right)
=(1/2,0),\ (1/2,1),\ (-1/2,-1),\ (-1/2,0).$

The differential cross-section can be written as
\begin{equation}
d\sigma \propto
\sum_{M_{i},M_{i}^{\prime}}\sum_{\lambda_{1},\lambda_{2}}\rho^{%
\Lambda_{b}}_{M_{i},M_{i}^{\prime}}|\mathcal{A}_{(\lambda _{1},\lambda
_{2})}(\Lambda _{b}\rightarrow \Lambda
V)|^{2}d^{1/2}_{M_{i}\lambda}d^{1/2}_{M_{i}^{\prime}\lambda}\text{exp}%
i(M_{i}^{\prime}-M_{i})\phi
\end{equation}
where we have taken into account the initial state helicity and have summed
over the final state helicities. The total angular momentum along the
helicity axis, $\lambda=M_{f}=\lambda_{1}-\lambda_{2}$, being fixed. As
parity is not conserved in the weak interactions therefore
\begin{equation*}
\mathcal{A}_{(\lambda _{1,}\lambda _{2})}(\Lambda _{b}\rightarrow \Lambda
V)\neq\mathcal{A}_{(-\lambda _{1},-\lambda _{2})}(\Lambda _{b}\rightarrow
\Lambda V)
\end{equation*}
It is worthwhile to introduce the helicity asymmetry parameter, $\alpha_{AS}$%
, for $\Lambda _{b}$ as:
\begin{equation}
\alpha_{AS}=\frac{\left\vert A_{\frac{1}{2},0}\right\vert ^{2}+\left\vert
A_{-\frac{1}{2},-1}\right\vert ^{2}-\left\vert A_{-\frac{1}{2},0}\right\vert
^{2}-\left\vert A_{\frac{1}{2},1}\right\vert ^{2}}{\left\vert A_{\frac{1}{2}
,0}\right\vert ^{2}+\left\vert A_{-\frac{1}{2},-1}\right\vert
^{2}+\left\vert A_{-\frac{1}{2},0}\right\vert ^{2}+\left\vert A_{\frac{1}{2}
,1}\right\vert ^{2}}
\end{equation}
The differential decay rate can be expressed in-terms of asymmetry parameter
as
\begin{equation}
\frac{d\sigma}{d\Omega}\propto 1+\alpha_{AS}\mathcal{P}^{\Lambda_{b}}\cos%
\theta+2\alpha_{AS}\Re(\rho^{\Lambda_{b}}_{+-}\text{exp}i\phi)\sin\theta
\end{equation}
Then, by averaging over the azimuthal angle, $\phi$, a standard relation is
obtained for the polar angular distribution:
\begin{equation}
\frac{d\sigma}{d\cos\theta}\propto 1+\alpha_{AS}\mathcal{P}%
^{\Lambda_{b}}\cos\theta
\end{equation}
where it can be noticed that polar angular dissymmetries are intimately
related to the initial polarization of the $\Lambda_{b}$ resonance.

\subsection{The Helicity Amplitude}

On the dynamical side, both tree and penguin diagrams are involved in the
evaluation of the Hadronic Matrix Elements($HME$). Heavy Quark effective
theory is extensively used for the calculation of $HME.$ \newline
In tree approximation, the effective interaction Hamiltonian, $\mathcal{H}%
^{eff}$ is,
\begin{equation}
\mathcal{H}^{eff}=\frac{G_{F}}{\sqrt{2}}V_{qb}V_{qs}^{\ast
}\sum_{i=1}^{10}C_{i}\left( m_{b}\right) O_{i}\left( m_{b}\right)
\end{equation}
where $C_{i}\left( m_{b}\right) $ are the Wilson Coefficients and the
operators, $O_{i}\left( m_{b}\right) $ can be understood as local operators
which govern the weak interaction of quarks in the given decay. They can be
written as
\begin{equation*}
O_{i}=\left( \bar{q}_{\alpha }\Gamma _{i1}q_{\beta }\right) \left( \bar{q}%
_{\mu }\Gamma _{i2}q_{\nu }\right)
\end{equation*}
where $\Gamma _{ij}$ denotes the gamma matrices.

By using the Factorization assumption one can get the helicity amplitude for
the decay $\Lambda _{b}\rightarrow \Lambda V$ $(1^{-})$ as
\begin{equation}
A_{\left( \lambda ,\lambda ^{\prime }\right) }=\frac{G_{F}}{\sqrt{2}}
f_{V}E_{V}\langle \Lambda (p^{\prime},s^{\prime})\left\vert \bar{s}\gamma
_{\mu }\left( 1-\gamma _{5}\right) b\right\vert \Lambda _{b}(p,s)\rangle
_{\left( \lambda ,\lambda ^{\prime }\right) }\left\{
V_{CKM}^{T}C_{i}^{T}-V_{CKM}^{P}C_{i}^{P}\right\}
\end{equation}
where $f_{V}$ and $E_{V}$ are the decay constant and energy of Vector meson.
$V_{CKM}^{T,P}=V_{qb}V_{qs}^{\ast }$ are the CKM matrix elements for the
tree and penguin diagrams while $C_{i}^{T,P}$ are Wilson Coefficients. The
baryonic matrix element $\mathcal{M}_{\left( \lambda ,\lambda ^{\prime
}\right) }^{\Lambda _{b}}\equiv \langle \Lambda
(p^{\prime},s^{\prime})\left\vert \bar{s}\gamma _{\mu }\left( 1-\gamma
_{5}\right) b\right\vert \Lambda _{b}(p,s)\rangle _{\left( \lambda ,\lambda
^{\prime }\right) }$ is calculated by using the Heavy Quark Effective
Theory(HQET), and read as
\begin{eqnarray}
\mathcal{M}_{\frac{1}{2},0}^{\Lambda _{b}} &=&-\frac{\left\vert \vec{P}%
_{V}\right\vert }{E_{V}}\left( \frac{m+m^{\prime }}{E_{\Lambda }+m^{\prime }
}\xi ^{-}\left( \omega \right) +2\xi _{2}\left( \omega \right) \right) \\
\mathcal{M}_{-\frac{1}{2},-1}^{\Lambda _{b}} &=&\frac{1}{\sqrt{2}}\left(
\frac{ \left\vert \vec{P}_{V}\right\vert }{E_{\Lambda }+m^{\prime}}\xi
^{-}\left( \omega \right) +\xi ^{+}\left( \omega \right) \right) \\
\mathcal{M}_{\frac{1}{2},1}^{\Lambda _{b}} &=&\frac{1}{\sqrt{2}}\left( \frac{
\left\vert \vec{P}_{V}\right\vert }{E_{\Lambda }+m^{\prime}}\xi ^{-}\left(
\omega \right) -\xi ^{+}\left( \omega \right) \right) \\
\mathcal{M}_{-\frac{1}{2},0}^{\Lambda _{b}} &=&\frac{\left\vert \vec{P}%
_{V}\right\vert ^{2}}{E_{V}\left( E_{V}+m^{\prime }\right) }\xi ^{-}\left(
\omega \right) +\xi ^{+}\left( \omega \right)
\end{eqnarray}
where $\left\vert \vec{P}_{V}\right\vert $ and $E_{V}$ are the momentum and
energy of vector meson in the rest frame of $\Lambda _{b}$, are given as
\begin{eqnarray}
\left\vert \vec{P}_{V}\right\vert &=&\frac{\sqrt{\left[ m^{2}-\left(
m_{V}+m^{\prime }\right) ^{2}\right] \left[ m^{2}-\left( m_{V}-m^{\prime
}\right) ^{2}\right] }}{2m} \\
E_{V} &=&\frac{m^{2}+m_{V}^{2}-m^{\prime2}}{2m},\ \text{and }E_{\Lambda }=%
\frac{m^{2}+m^{\prime 2}-m_{V}^{2}}{2m}
\end{eqnarray}
and the form factors $\xi ^{\pm }\left( \omega \right) =\xi _{1}\left(
\omega \right) \pm \xi _{2}\left( \omega \right) $ are defined for
convenience. While the form factors $\xi _{1,2}\left( \omega \right) $ are
evaluated in terms of heavy quark effective form factors $F_{i}^{\prime }s$
as
\begin{eqnarray}
\xi _{1}\left( \omega \right) &=&\frac{1}{2}\left[ 2F_{1}\left( \omega
\right) +F_{2}\left( \omega \right) +F_{3}\left( \omega \right) \left( 1+%
\frac{m_{\Lambda _{b}}}{m_{\Lambda }}\right) \right] \\
\xi _{2}\left( \omega \right) &=&\frac{1}{2}F_{2}\left( \omega \right)
\end{eqnarray}

\subsection{Polarizations and Angular Distributions}

Parity violation in $\Lambda _{b}$ weak decays into $\Lambda$, $V$
necessarily leads to a polarization process of the two intermediate
resonances $\Lambda $ and $V$. In order to determine the vector-polarization
of each resonance, a new set of axis is defined as
\begin{equation*}
\vec{e}_{L}=\frac{\vec{p}}{\left\vert \vec{p}\right\vert };\ \ \ \ \vec{e}%
_{z}=\vec{n}=\frac{\vec{p}_{p}\times \vec{p}_{b}}{\left\vert \vec{p}
_{p}\times \vec{p}_{b}\right\vert };\ \ \ \ \vec{e}_{N}=\vec{e}_{z}\times
\vec{e}_{L};\ \ \ \ \vec{e}_{T}=\vec{e}_{L}\times \vec{e}_{N}
\end{equation*}
where $\vec{p}$ is the momentum of $\Lambda$ and $\vec{n}$ is the
quantization plane as defined in previous section.

In this new frame, the vector-polarization of any resonance defined in the
original $\Lambda _{b}$ frame can be written as:
\begin{equation*}
\vec{\mathcal{P}}^{i}=P_{L}\vec{e}_{L}+P_{T}\vec{e}_{T}+P_{N}\vec{e}_{N}
\end{equation*}
where $i=\Lambda $ or $V$ and $P_{L},P_{N},P_{T}$ are longitudinal, normal
and transverse polarizations of the decay resonance.

It is worth noticing that the basis vectors $\vec{e}_{L}$, $\vec{e}_{N}$ and
$\vec{e}_{T}$ have the following properties according to parity and TR:
\textit{P-odd,T-odd; P-odd,T-odd} and \textit{P-even, T-even} respectively,
while the polarization-vector $\vec{\mathcal{P}}$ is P-even and T-odd. So
using these properties we can get $P_{L}=P-odd,T-even$, $P_{N}=P-odd,T-even$
and $\mathbf{P}_{T}=P-even,\mathbf{T-odd}$. As $\mathbf{P}_{T}$ is T-odd so
any non-zero value of this polarization will be a clear signature of Time
Reversal violation.

\subsection{Polarization of final state resonances}

Intermediate resonance states, $\Lambda $ and $V$ , can be described by a
density-matrix named $\rho ^{f}$ whose analytic expression is given by
standard quantum-mechanical relations:
\begin{equation}
\rho ^{f}=\mathcal{T}^{\dagger }\rho ^{\Lambda _{b}}\mathcal{T}  \label{4p1}
\end{equation}
where $\mathcal{T}$ is the transition-matrix related to the S-matrix by $%
S=1+i\mathcal{T}$ . The matrix elements of the SDM $\rho^{f}$ are obtained
from $\eqref{4p1}$ by projecting the operators involved in that expression
onto the initial and final states. The latter ones are characterized by a
given three-momentum in the $\Lambda_{b}$ center-of-mass system and by a
pair of helicities, $\lambda_{1}$ and $\lambda_{2}$, corresponding to each
resonance $\Lambda$ and $V$. Therefore the SDM of this two-particle system
is endowed with two pairs of indices, as:
\begin{eqnarray}
\rho _{\lambda_{1} \lambda_{1} ^{\prime }\lambda_{2} \lambda_{2} ^{\prime
}}^{f} &=&\sum_{M,M^\prime}F^{JM}_{\lambda_{1} \lambda_{2}}(\theta,\phi)\rho
_{M,M^{\prime }}^{\Lambda_{b}}F^{JM^{\prime *}}_{\lambda_{1}^{\prime }
\lambda_{2}^{\prime }}(\theta,\phi)  \label{4p2} \\
F^{JM}_{\lambda_{1} \lambda_{2}}(\theta,\phi)
&=&\langle\theta,\phi;\lambda_{1}\lambda_{2}|\mathcal{T}|JM\rangle
\label{4p3}
\end{eqnarray}
where $\theta$ and $\phi$ are the polar and azimuthal angles of the momentum
of $\Lambda$ resonance in the $\Lambda_{b}$ rest frame respectively as shown
in fig (4.1).

Taking into account the angular momentum conservation, we can write%
\begin{equation*}
\chi=\lambda_{1}-\lambda_{2}\ \ \ \ \
\chi^{\prime}=\lambda_{1}^{\prime}-\lambda_{2}^{\prime}
\end{equation*}
where $\chi, \ \chi^{\prime}=\pm1/2$. So one can write the eq.$\eqref{4p2}$
of the SDM as
\begin{equation}
\rho _{\lambda_{1} \lambda_{1} ^{\prime }\chi\chi^{\prime }}^{f}
=\sum_{M,M^\prime}F^{JM}_{\lambda_{1} \lambda_{1}-\chi}(\theta,\phi)\rho
_{M,M^{\prime }}^{\Lambda_{b}}F^{JM^{\prime *}}_{\lambda_{1}^{\prime }
\lambda_{1}^{\prime }-\chi^{\prime}}(\theta,\phi)  \label{4p4} \\
\end{equation}
Jacob-Wick helicity formalism gives
\begin{equation}
F^{JM}_{\lambda_{1} \lambda_{2}}(\theta,\phi)=N_{J}A^{J}_{\lambda_{1}
\lambda_{2}}D^{J*}_{M,\chi^{\prime}}(\theta,\phi,0)
\end{equation}
where
\begin{equation}
A_{\lambda_{1} ,\lambda_{2} }^{J}=4\pi \left( \frac{m_{b}}{\left\vert \vec{p}%
_{b}\right\vert }\right) \langle J,M;\lambda_{1} ,\lambda_{2} \left\vert
\mathcal{T}\right\vert J,M\rangle
\end{equation}
is the helicity amplitude, which is evaluated in the previous section by
using the HQET.

After summing over the initial polarizations of $M,\ M^{\prime }$ of $%
\Lambda _{b}$, and taking into account the angular momentum conservation and
properties of Wigner matrices, the SDM of the final states will be
\begin{equation}
\rho _{\lambda \lambda ^{\prime }\chi \chi ^{\prime }}^{f}=\frac{1}{4\pi }
\left\{ A_{\lambda ,\lambda -\chi }A_{\lambda ^{\prime },\lambda ^{\prime
}-\chi }^{\ast }\left( 1+4\chi P_{1}^{\Lambda _{b}}\right) \delta _{\chi
,\chi ^{\prime }} \\
+2A_{\lambda ,\lambda ^{\prime }-\chi }A_{\lambda ,\lambda ^{\prime }+\chi
}^{\ast }\left( P_{2}^{\Lambda _{b}}+2i\chi P_{3}^{\Lambda _{b}}\right)
\delta _{\chi ,-\chi ^{\prime }} \right\}  \label{4p5}
\end{equation}
where we have draped the index $J$ from the amplitude and the index 1 from
helicities. By angular momentum conservation, $\chi^{\prime }s$ have the
form,
\begin{equation*}
\chi =\lambda _{1}-\lambda _{2}=\pm \frac{1}{2};\ \ \ \chi ^{\prime
}=\lambda _{1}^{\prime }-\lambda _{2}^{\prime }=\pm \frac{1}{2}
\end{equation*}
Moreover we have set
\begin{equation}
P_{1}^{\Lambda_{b}}=\vec{\mathcal{P}}^{\Lambda_{b}} \cdot \hat{p}, \ \ \
P_{2}^{\Lambda_{b}}=\vec{\mathcal{P}}^{\Lambda_{b}}\cdot\vec{e}_{N}, \ \ \
P_{3}^{\Lambda_{b}}=\vec{\mathcal{P}}^{\Lambda_{b}}\cdot\hat{r},  \label{4p8}
\end{equation}
where $\hat{p}$ is the unit vector of the $\Lambda$ momentum and
\begin{equation*}
\vec{e}_{N}=\vec{e}_{T}\times\hat{p},\ \ \ \ \vec{e}_{T}=\frac{\vec{n}\times%
\hat{p}}{|\vec{n}\times\hat{p}|}, \ \ \ \ \hat{r}=\vec{e}_{T}\times\vec{n}.
\end{equation*}

The angular distribution of the decay products, $W(\theta ,\phi )$, can be
deduced from the SDM, according to the formulae
\begin{equation*}
W\left( \theta ,\phi \right) =Tr\left( \rho _{\lambda \lambda ^{\prime }\chi
\chi ^{\prime }}^{f}\right)
\end{equation*}
Taking into account $\rho _{\lambda \lambda ^{\prime }\chi \chi ^{\prime
}}^{f}$, we get
\begin{equation*}
W\left( \theta ,\phi \right) =\frac{1}{4\pi }\left( G_{W}+\Delta
G_{W}P_{1}^{\Lambda _{b}}\right)
\end{equation*}
where
\begin{eqnarray}
G_{W} &=&\left\vert A_{\frac{1}{2},0}\right\vert ^{2}+\left\vert A_{-\frac{1%
}{2},-1}\right\vert ^{2}+\left\vert A_{-\frac{1}{2},0}\right\vert
^{2}+\left\vert A_{\frac{1}{2},1}\right\vert ^{2} \\
\Delta G_{W} &=&2\left( \left\vert A_{\frac{1}{2},0}\right\vert
^{2}+\left\vert A_{-\frac{1}{2},-1}\right\vert ^{2}-\left\vert A_{-\frac{1}{%
2 },0}\right\vert ^{2}-\left\vert A_{\frac{1}{2},1}\right\vert ^{2}\right)
\end{eqnarray}

The polarization vectors of the resonance states can be evaluated as
\begin{equation}
\vec{\mathcal{P}}^{i}=\frac{Tr\left( \rho _{\lambda \lambda ^{\prime }\chi
\chi ^{\prime }}^{f}\cdot \vec{s}^{i}\right) }{Tr\left( \rho _{\lambda
\lambda ^{\prime }\chi \chi ^{\prime }}^{f}\right) }=\frac{Tr\left( \rho
_{\lambda \lambda ^{\prime }\chi \chi ^{\prime }}^{f}\cdot \vec{s}%
^{i}\right) }{ W\left( \theta ,\phi \right) }  \label{4p6}
\end{equation}
so that,
\begin{equation}
W\left( \theta ,\phi \right) \vec{P}^{i}=Tr\left( \rho _{\lambda \lambda
^{\prime }\chi \chi ^{\prime }}^{f}\cdot \vec{s}^{i}\right)  \label{4p7}
\end{equation}

where $\vec{s}\equiv (s_{x},s_{y},s_{z})$ denotes the spin vector operator
of the resonance state.

\subsubsection{Polarization Vector of $\Lambda$}

The components of the polarization vector of $\Lambda $ can be evaluated
from the above relations by summing over the helicity states. The three
components of the $\mathcal{P}^{\Lambda}$ have the following form:
\begin{eqnarray*}
W\left( \theta ,\phi \right) P_{L}^{\Lambda }\left( \theta ,\phi \right)
&\propto& \gamma(+1/2)\left( |\mathcal{A}_{1/2,0}|^{2}-|\mathcal{A}%
_{-1/2,-1}|^{2}\right)+\gamma(-1/2)\left( |\mathcal{A}_{1/2,1}|^{2}-|%
\mathcal{A}_{-1/2,0}|^{2}\right) \\
W\left( \theta ,\phi \right) P_{N}^{\Lambda }\left( \theta ,\phi \right)
&\propto& \Re\left( \mathcal{A}_{1/2,0}\mathcal{A}_{-1/2,0}^{*}-\mathcal{P}%
^{\Lambda_{b}}\sin\theta+2\Re(e^{i\phi}\rho^{\Lambda_{b}}_{+-})\cos\theta+2i%
\Im(e^{i\phi}\rho^{\Lambda_{b}}_{+-}) \right) \\
W\left( \theta ,\phi \right) P_{T}^{\Lambda }\left( \theta ,\phi \right)
&\propto& -\Im\left( \mathcal{A}_{1/2,0}\mathcal{A}_{-1/2,0}^{*}-\mathcal{P}%
^{\Lambda_{b}}\sin\theta+2\Re(e^{i\phi}\rho^{\Lambda_{b}}_{+-})\cos\theta+2i%
\Im(e^{i\phi}\rho^{\Lambda_{b}}_{+-}) \right)
\end{eqnarray*}
where,
\begin{equation*}
\gamma(\pm1/2)=\frac{1}{2}\left( 1\pm \mathcal{P}^{\Lambda_{b}} \cos\theta
\pm 2\Re(e^{i\phi}\rho^{\Lambda_{b}}_{+-})\sin\theta\right)
\end{equation*}

One can get the explicit relations for the components of polarization vector
of intermediate states, which only depends on the helicity amplitude, as:
\begin{eqnarray}
W\left( \theta ,\phi \right) P_{L}^{\Lambda }\left( \theta ,\phi \right) &=&
\frac{1}{4\pi }\left( G_{L}^{\Lambda }+\Delta G_{L}^{\Lambda }P_{1}^{\Lambda
_{b}}\right) \\
W\left( \theta ,\phi \right) P_{T}^{\Lambda }\left( \theta ,\phi \right) &=&
\frac{1}{4\pi }\left( G_{T}^{\Lambda }P_{2}^{\Lambda _{b}}+\Delta
G_{T}^{\Lambda }P_{3}^{\Lambda _{b}}\right) \\
W\left( \theta ,\phi \right) P_{N}^{\Lambda }\left( \theta ,\phi \right) &=&
\frac{1}{4\pi }\left( G_{N}^{\Lambda }P_{2}^{\Lambda _{b}}+\Delta
G_{N}^{\Lambda }P_{3}^{\Lambda _{b}}\right)
\end{eqnarray}
where%
\begin{eqnarray}
G_{L}^{\Lambda } &=&\frac{1}{2}\left( \left\vert A_{\frac{1}{2}%
,0}\right\vert ^{2}-\left\vert A_{-\frac{1}{2},-1}\right\vert
^{2}-\left\vert A_{-\frac{1}{2},0}\right\vert ^{2}+\left\vert A_{\frac{1}{2}
,1}\right\vert ^{2}\right) \\
\Delta G_{L}^{\Lambda } &=&\left\vert A_{\frac{1}{2},0}\right\vert
^{2}-\left\vert A_{-\frac{1}{2},-1}\right\vert ^{2}+\left\vert A_{-\frac{1}{%
2 },0}\right\vert ^{2}-\left\vert A_{\frac{1}{2},1}\right\vert ^{2} \\
G_{T}^{\Lambda } &=&-2\Im \left( A_{\frac{1}{2},0}A_{-\frac{1}{2},0}^{\ast
}+A_{\frac{1}{2},1}A_{-\frac{1}{2},-1}^{\ast }\right) \\
\Delta G_{T}^{\Lambda } &=&2\Re \left( A_{\frac{1}{2},0}A_{-\frac{1}{2}
,0}^{\ast }-A_{\frac{1}{2},1}A_{-\frac{1}{2},-1}^{\ast }\right) \\
G_{N}^{\Lambda } &=&2\Re \left( A_{\frac{1}{2},0}A_{-\frac{1}{2},0}^{\ast
}+A_{\frac{1}{2},1}A_{-\frac{1}{2},-1}^{\ast }\right) \\
\Delta G_{N}^{\Lambda } &=&-2\Im \left( A_{\frac{1}{2},0}A_{-\frac{1}{2}
,0}^{\ast }-A_{\frac{1}{2},1}A_{-\frac{1}{2},-1}^{\ast }\right)
\end{eqnarray}

\subsubsection{Polarization Vector of $V(1^-)$}

To calculate the components of the polarization of Vector meson we have to
take into account its spin $\vec{s}$ vector and corresponding helicity
states, one can get:
\begin{eqnarray}
W\left( \theta ,\phi \right) P_{L}^{\Lambda V}\left( \theta ,\phi \right) &=&%
\frac{1}{4\pi }\left( G_{L}^{V}+\Delta G_{L}^{V}P_{1}^{\Lambda _{b}}\right)
\\
W\left( \theta ,\phi \right) P_{T}^{V}\left( \theta ,\phi \right) &=&\frac{1
}{4\pi }\left( G_{T}^{V}P_{2}^{\Lambda _{b}}+\Delta G_{T}^{V}P_{3}^{\Lambda
_{b}}\right) \\
W\left( \theta ,\phi \right) P_{N}^{V}\left( \theta ,\phi \right) &=&\frac{1
}{4\pi }\left( \Delta G_{T}^{V}P_{2}^{\Lambda _{b}}-G_{T}^{V}P_{3}^{\Lambda
_{b}}\right)
\end{eqnarray}
where%
\begin{eqnarray}
G_{L}^{\Lambda } &=&-2\left( \left\vert A_{-\frac{1}{2},-1}\right\vert
^{2}+\left\vert A_{\frac{1}{2},1}\right\vert ^{2}\right) \\
\Delta G_{L}^{\Lambda } &=&\left\vert A_{-\frac{1}{2},-1}\right\vert
^{2}-\left\vert A_{\frac{1}{2},1}\right\vert ^{2} \\
G_{T}^{\Lambda } &=&-2\sqrt{2}\Im \left( A_{\frac{1}{2},1}A_{\frac{1}{2}
,0}^{\ast }-A_{-\frac{1}{2},-1}A_{-\frac{1}{2},0}^{\ast }\right) \\
\Delta G_{T}^{\Lambda } &=&2\sqrt{2}\Re \left( A_{\frac{1}{2},1}A_{\frac{1}{%
2 },0}^{\ast }+A_{-\frac{1}{2},-1}A_{-\frac{1}{2},0}^{\ast }\right)
\end{eqnarray}

\subsubsection{Polarization Correlations}

We now define four polarization correlations, similar to those defined by
Chiang and Wolfenstein$\cite{cw}$:
\begin{eqnarray}
W\left( \theta ,\phi \right) P_{TT(NN)}(\theta\phi)&=&\frac{1}{2 }Tr\left[
\rho^{f}\sigma^{\Lambda}_{y(x)}s^{V}_{y(x)}\right]  \label{4pc1} \\
W\left( \theta ,\phi \right) P_{TN(NT)}(\theta\phi)&=&\frac{1}{2 }Tr\left[
\rho^{f}\sigma^{\Lambda}_{y(x)}s^{V}_{x(y)}\right]  \label{4pc2}
\end{eqnarray}
These observables are related to the angular distributions of the decay
products of resonances $\Lambda$ and $V$, similar to those considered in
ref. $\cite{datta}$.

Making use of eq.$\eqref{4p5}$ in the above equation, we get
\begin{eqnarray}
W\left( \theta ,\phi \right) P_{TT}(\theta\phi)&=&\frac{1}{4\pi }\left(
G_{TT}+\Delta G_{TT}P_{1}^{\Lambda_{b}}\right)  \label{4pc3} \\
W\left( \theta ,\phi \right) P_{NT}(\theta\phi)&=&\frac{1}{4\pi }\left(
\Delta G_{TN}+ G_{TN}P_{1}^{\Lambda_{b}}\right)  \label{4pc4} \\
W\left( \theta ,\phi \right) P_{TN}(\theta\phi)&=&\frac{1}{4\pi }\left(
G_{TN}+\Delta G_{TN}P_{1}^{\Lambda_{b}}\right)  \label{4pc5} \\
W\left( \theta ,\phi \right) P_{NN}(\theta\phi)&=&-\frac{1}{4\pi }\left(
G_{TT}+\Delta G_{TT}P_{1}^{\Lambda_{b}}\right)  \label{4pc6}
\end{eqnarray}
with
\begin{eqnarray*}
G_{TT} &=&-\frac{1}{\sqrt2}\Re\left( A_{-\frac{1}{2},-1}A_{\frac{1}{2}%
,0}^{*}+ A_{\frac{1}{2},1}A_{-\frac{1}{2},0}^{*}\right), \\
\Delta G_{TT} &=&-\sqrt2\Re\left( A_{-\frac{1}{2},-1}A_{\frac{1}{2},0}^{*}-
A_{\frac{1}{2},1}A_{-\frac{1}{2},0}^{*}\right), \\
G_{TN} &=&\sqrt2\Im\left( A_{-\frac{1}{2},-1}A_{\frac{1}{2},0}^{*}+ A_{\frac{%
1}{2},1}A_{-\frac{1}{2},0}^{*}\right), \\
\Delta G_{TN} &=& \frac{1}{\sqrt2}\Im\left( A_{-\frac{1}{2},-1}A_{\frac{1}{2}%
,0}^{*}- A_{\frac{1}{2},1}A_{-\frac{1}{2},0}^{*}\right)
\end{eqnarray*}

\subsection{Parametrization of Observables}

We can write a model independent parametrization, based on the previous
formulae, of the angular distribution, of the polarization of $\Lambda $, $V$
and polarization correlations. Our purpose for parameterizations is to
describe the observables in terms of a minimum number of independent
parameters.

The formulae of the angular distribution and of the polarization of $\Lambda$
can be rewritten as
\begin{eqnarray}
W\left( \theta ,\phi \right) &=&\frac{1}{4\pi }G_{W}\left( 1+2\alpha
_{W}P_{1}^{\Lambda _{b}}\right)  \label{4po1} \\
\vec{\mathcal{P}}^{\Lambda } &=&\frac{1}{1+2\alpha _{W}P_{1}^{\Lambda _{b}}}%
\left( C_{L}\vec{e}_{L}+C_{T}\vec{e}_{T}+C_{N}\vec{e}_{N}\right)
\label{4po2}
\end{eqnarray}
with
\begin{eqnarray}
\alpha _{W} &=&\frac{\Delta G_{W}}{2G_{W}} \\
C_{L} &=&B_{L}\left( 1+2\alpha _{L}P_{1}^{\Lambda _{b}}\right) \\
C_{T} &=&B_{T}\left( P_{2}^{\Lambda _{b}}+2\alpha _{T}P_{3}^{\Lambda
_{b}}\right) \\
C_{N} &=&B_{N}\left( P_{2}^{\Lambda _{b}}+2\alpha _{N}P_{3}^{\Lambda
_{b}}\right)
\end{eqnarray}
where
\begin{eqnarray}
B_{L} &=&\frac{G_{L}^{\Lambda }}{G_{W}};\ \ \ B_{T}=\frac{G_{T}^{\Lambda }}{
G_{W}};\ \ \ B_{N}=\frac{G_{N}^{\Lambda }}{G_{W}}; \\
\alpha _{L} &=&\frac{\Delta G_{L}^{\Lambda }}{2G_{L}^{\Lambda }};\ \ \alpha
_{T}=\frac{\Delta G_{T}^{\Lambda }}{2G_{T}^{\Lambda }};\ \ \alpha _{N}=\frac{%
\Delta G_{N}^{\Lambda }}{2G_{N}^{\Lambda }}
\end{eqnarray}

One can get similar correlations for Vector meson. Where as the polarization
correlations can be reshaped as:
\begin{eqnarray}
P_{TT}&=&\frac{1}{1+2\alpha _{W}P_{1}^{\Lambda _{b}}}B_{TT}\left(
1+2P_{1}^{\Lambda_{b}}\alpha_{TT}\right) \\
P_{TN}&=&\frac{1}{1+2\alpha _{W}P_{1}^{\Lambda _{b}}}B_{TN}\left(
1+2P_{1}^{\Lambda_{b}}\alpha_{TN}\right)
\end{eqnarray}
where,
\begin{eqnarray}
B_{TT} &=&\frac{G_{TT}}{G_{W}};\ \ \ B_{TN}=\frac{G_{TN}}{G_{W}}; \\
\alpha _{TT} &=&\frac{\Delta G_{TT}}{2G_{TT}};\ \ \alpha _{TN}=\frac{\Delta
G_{TN}}{2G_{TN}}.
\end{eqnarray}

\subsection{TRV, CPV and CPT Tests}

Now we illustrate properties of the above observables under discrete
transformations and indicate the violation of parameters introduced in the
previous section under discrete transformations.\newline

\subsubsection{Time reversal violation}

Since the helicity is invariant under the time reversal (TR) operation but
the rotationally invariant helicity amplitudes transform under time TR in
such a way that
\begin{equation}
A_{\lambda ,\lambda ^{\prime }}A_{-\lambda ,-\lambda ^{\prime }}^{\ast
}\longrightarrow A_{\lambda ,\lambda ^{\prime }}^{\ast }A_{-\lambda
,-\lambda ^{\prime }}  \label{cpt1}
\end{equation}
This follows from the anti-unitarity character of TR. As the parameters $%
G_{T}^{\Lambda}$, $\Delta G_{T}^{\Lambda}$, $G_{T}^{V}$, $\Delta G_{T}^{V}$,
$G_{TN}$ and $\Delta G_{TN}$ reverse sign in TR operation, and as such these
parameters along with eq.$\eqref{4p8}$ suggest that the transverse
polarizations $P_{T}^{\Lambda }\left( \theta ,\phi \right) $, $\
P_{T}^{V}\left( \theta ,\phi \right) $ and the polarization correlations $%
P_{NT}(\theta,\phi)$ and $P_{TN}(\theta,\phi)$ are T-odd under this
transformation. So non-zero value of any of these observables will be a
clear signature of direct TRV. Our numerical results are discussed in the
next section for TRV. These observable are also promising for possible
effects of New Physics as discussed in $\cite{datta1}$ and $\cite{datta2}.$

\subsubsection{CP-Violation}

The CP transformation causes, according to the usual phase conventions$\cite%
{rdb}$
\begin{equation}
A_{\lambda ,\lambda ^{\prime }}\rightarrow -\bar{A}_{-\lambda ,-\lambda
^{\prime }}  \label{cpt2}
\end{equation}
where the barred amplitude refers to the $\bar{\Lambda}_{b}$ decay. For
detecting possible CP violation we define the following asymmetry
parameters, which depend on the observables defined in the previous section,
as:
\begin{eqnarray}
R_{W} &=&\frac{G_{W}-\bar{G}_{W}}{G_{W}+\bar{G}_{W}},\ \ \ \ \ \ \ R_{L}=%
\frac{ B_{L}+\bar{B}_{L}}{B_{L}-\bar{B}_{L}} \\
R_{T} &=&\frac{B_{T}+\bar{B}_{T}}{B_{T}-\bar{B}_{T}},\ \ \ \ \ \ \ \ \
R_{N}= \frac{B_{N}-\bar{B}_{N}}{B_{N}-\bar{B}_{N}} \\
\gamma _{W} &=&\frac{\alpha _{W}+\bar{\alpha}_{W}}{\alpha _{W}-\bar{\alpha}
_{W}},\ \ \ \ \ \ \ \ \ \gamma _{L}=\frac{\alpha _{L}+\bar{\alpha}_{L}}{%
\alpha _{L}-\bar{\alpha}_{L}} \\
\gamma _{T} &=&\frac{\alpha _{T}+\bar{\alpha}_{T}}{\alpha _{T}-\bar{\alpha}
_{T}},\ \ \ \ \ \ \ \ \ \ \gamma _{N}=\frac{\alpha _{N}+\bar{\alpha}_{N}}{%
\alpha _{N}-\bar{\alpha}_{N}} \\
R_{TT}&=& \frac{B_{TT}-\bar{B}_{TT}}{B_{TT}+\bar{B}_{TT}}, \ \ \ \ R_{TN}=
\frac{B_{TN}+\bar{B}_{TN}}{B_{TN}-\bar{B}_{TN}} \\
\gamma_{TT}&=& \frac{\alpha_{TT}+\bar{\alpha}_{TT}}{\alpha_{TT}-\bar{\alpha}%
_{TT}}, \ \ \ \ \ \gamma_{TN}=\frac{\alpha_{TN}+\bar{\alpha}_{TN}}{%
\alpha_{TN}-\bar{\alpha}_{TN}}
\end{eqnarray}
Any non-zero value of the above ratios would be a CPV asymmetry parameter.
The numerical values corresponding to these observables are discussed in the
next chapter. It is interesting to see that all the above ratios are even
under time reversal, therefore they can be applied to test for the CPT
theorem and possibly the signature of New Physics.

\section{Physical Results and Conclusions}

In this chapter we will put our numerical results of the observables, which
we have analytically formulated in the previous chapters.

\subsection{Transition form factors and Branching Ratios}

The constituent quark masses are used in order to calculate the electroweak
form factor transitions between baryons and our used values are:
\begin{table}[ht]
\caption{Quark masses in GeV}\centering
\begin{tabular}{ccccc}
\hline\hline
$m_{u}$ & $m_{d}$ & $m_{s}$ & $m_{c}$ & $m_{b}$ \\ \hline
0.350 & 0.350 & 0.500 & 1.300 & 4.900 \\ \hline\hline
\end{tabular}%
\end{table}

For hadron masses, we shall use the following values:
\begin{table}[ht]
\caption{Hadron masses in GeV}\centering
\begin{tabular}{ccccc}
\hline\hline
$m_{\Lambda_{b}}$ & $m_{\Lambda}$ & $m_{J/\psi}$ & $m_{\rho}$ & $m_{\omega}$
\\ \hline
5.624 & 1.115 & 3.096 & 0.769 & 0.782 \\ \hline\hline
\end{tabular}%
\end{table}

The baryon heavy-to-light form factors, $F_{i}(\omega)$ and $G_{i}(\omega)$,
depending on the inner structure of the hadrons have been calculated in
Chapter 3. The decay constants for vector mesons, $f_{V}$, do not suffer
from uncertainties as large as those for form factors since they are well
determined experimentally from leptonic and semi-leptonic decays. Let us
first recall the usual definition for a vector meson,
\begin{equation}
c\langle V(q)|\bar{q_1}\gamma_{\mu}q_2|0\rangle=f_V m_V \epsilon_{\mu}
\end{equation}
where $m_V$ and $\epsilon_{\mu}$ are respectively the mass and polarization
4-vector of the vector meson, and c is a constant depending on the given
meson for example: $c=\sqrt{2}$ for the $\rho^0$ and $\omega^0$ and $c = 1$
for $J/\psi$. Numerically, in our calculations, for the decay constants we
take (in MeV), $f_{\rho} = 209$ , $f_{\omega} = 187$ , $f_{J/\psi} = 400$.
Finally, for the total $\Lambda_b$ decay width, $\Gamma_{\Lambda_{b}}=\frac{1%
}{\tau_{\Lambda_{b}}}$, we use $\tau_{\Lambda_{b}}= 1.229\pm 0.080 ps$.

We have calculated the transition form factors for the decay $\Lambda
_{b}\rightarrow \Lambda V,$ by using the heavy quark symmetry. The baryonic
form factors involved in evaluation of transition matrix $\mathcal{M}%
_{\left( \lambda ,\lambda ^{\prime }\right) }^{\Lambda _{b}}\equiv \langle
\Lambda (p^{\prime},s^{\prime})\left\vert \bar{s}\gamma _{\mu }\left(
1-\gamma _{5}\right) b\right\vert \Lambda _{b}(p,s)\rangle _{\left( \lambda
,\lambda ^{\prime }\right) }$ and calculated in chapter 3 are plotted verses
the invariant velocity for $\omega$ in Figures [5.1], [5.2] and [5.3].

\begin{figure}[ht]
\centering
\includegraphics[scale=0.3]{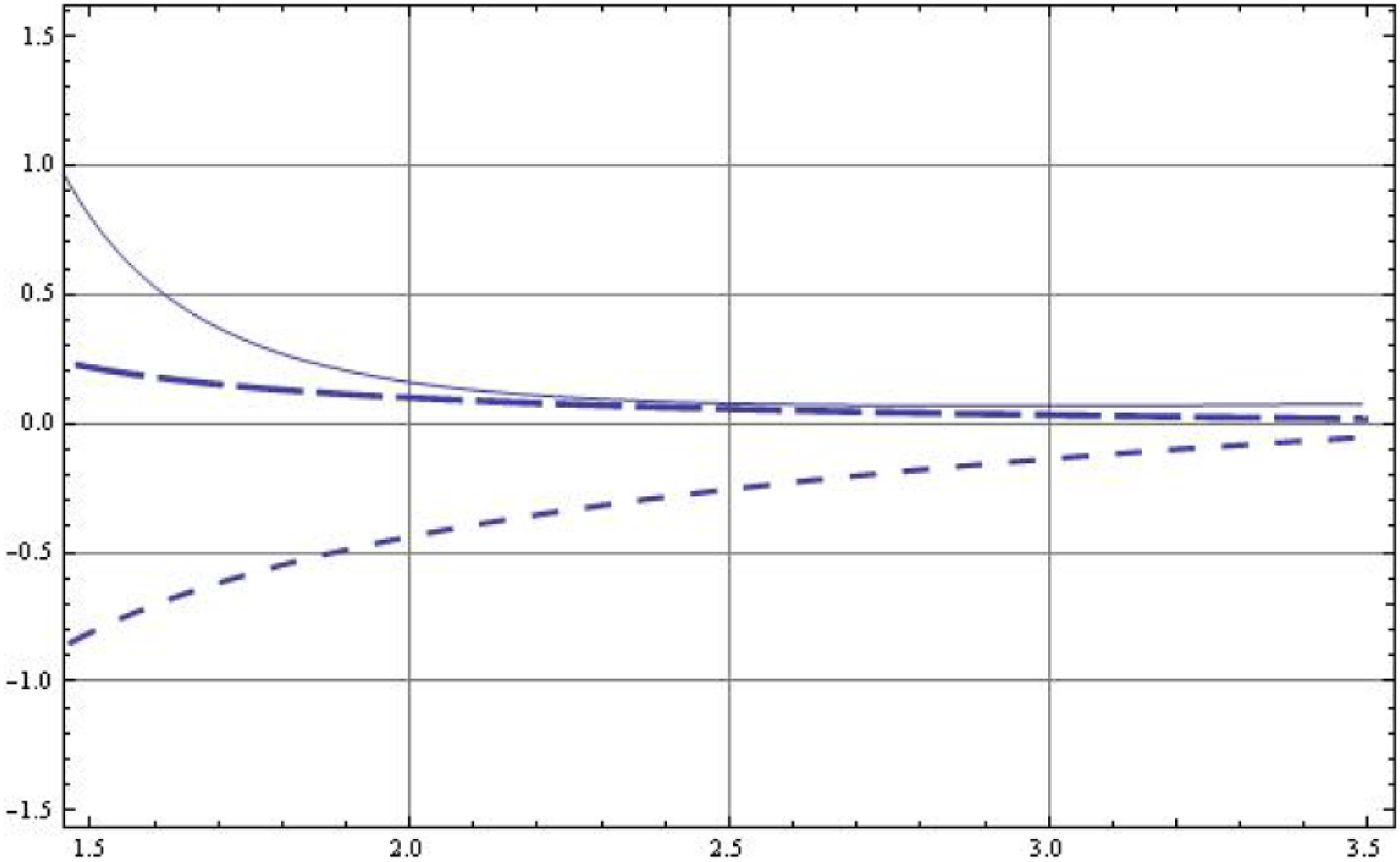} 
\caption{Farm Factors $F_{1}$ solid curve, $F_{2}$ short-dashed curve and $%
F_{3}$ long-dashed curve verses invariant velocity transfer, $\protect\omega$%
}
\end{figure}
\begin{figure}[ht]
\centering
\includegraphics[scale=0.3]{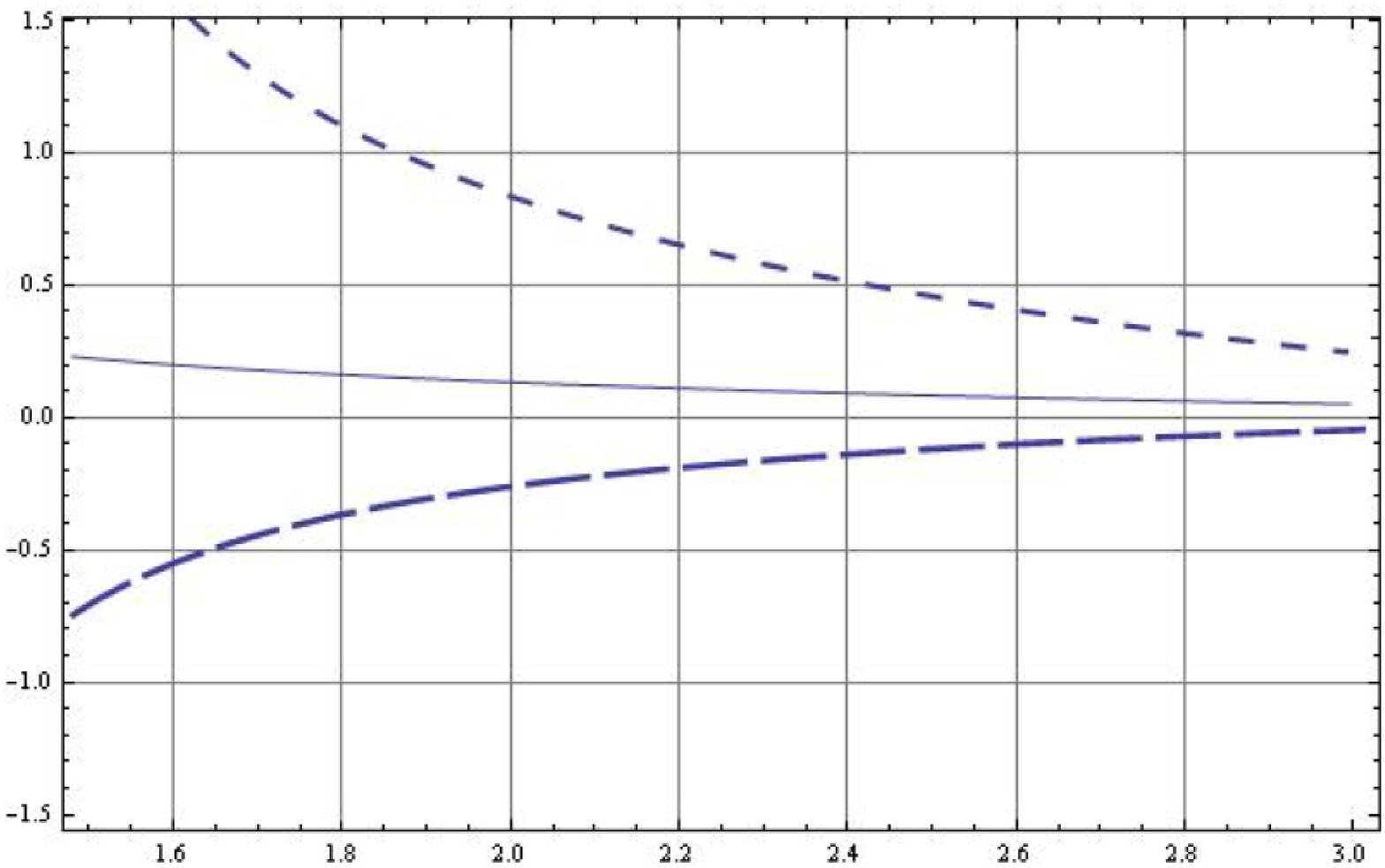} 
\caption{Farm Factors $G_{1}$ solid curve, $G_{2}$ short-dashed curve and $%
G_{3}$ long-dashed curve verses invariant velocity transfer, $\protect\omega$%
}
\end{figure}
\begin{figure}[ht]
\centering
\includegraphics[scale=0.3]{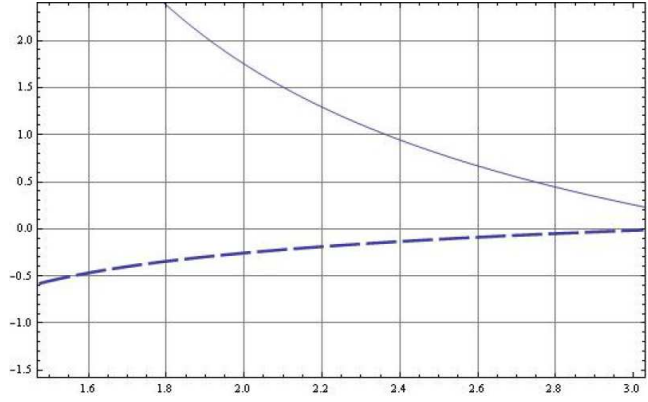} 
\caption{Farm Factors $\protect\xi_{1}$ solid curve, $\protect\xi_{2}$
dashed curve verses invariant velocity transfer, $\protect\omega$}
\end{figure}

The kinematical analysis and the factorization procedure developed in
chapters 3 and 4, allows us to compute the branching ratios of $%
\Lambda_{b}\to \Lambda J/\psi$, $\Lambda_{b}\to \Lambda \rho^{0}$ and $%
\Lambda_{b}\to \Lambda \omega$. We have not taken into account the
non-factorizable effects coming from the color octet contribution,
calculations have been performed by restricting the number of colors, $N_c$
takes the value 3.

The decay width of any process like $\Lambda_{b} \to \Lambda V$ is given by
the following formula $\cite{Pakv}$,
\begin{equation}
\Gamma(\Lambda_{b}\to \Lambda V)=\left( \frac{E_{\Lambda}+m_{\Lambda}}{%
m_{\Lambda_{b}}}\right)\frac{P_V}{16\pi^2}\int_{\Omega}\vert\sum_{\lambda_{%
\Lambda},\lambda_{V}}\mathcal{A}_{\lambda_{\Lambda},\lambda_{V}}(\Lambda_{b}%
\to \Lambda V)\vert^{2}d\Omega  \label{5br}
\end{equation}%
where in the $\Lambda_{b}$ rest frame, momentum of vector meson is,
\begin{equation}
\left\vert \vec{P}_{V}\right\vert =\frac{\sqrt{\left[m^{2}-\left(
m_{V}+m_{\prime }\right) ^{2}\right] \left[m^{2}-\left( m_{V}-m_{\prime
}\right) ^{2}\right] }}{2m}
\end{equation}
In eq.$\eqref{5br}$, $E_{\Lambda}$ is the energy of $\Lambda$ baryon and $%
\Omega$ is the decay solid angle. The helicity amplitude $\mathcal{A}%
_{\lambda_{\Lambda},\lambda_{V}}(\Lambda_{b}\to \Lambda V)$ is calculated in
chapter 4. Branching ratios, $\mathcal{BR}$, have been calculated for the
above three decays their values are,
\begin{eqnarray}
\mathcal{BR}(\Lambda_{b}\to \Lambda J/\psi)&=&6.3\times 10^{-4} \\
\mathcal{BR}(\Lambda_{b}\to \Lambda \rho^{0})&=&3.8\times 10^{-6} \\
\mathcal{BR}(\Lambda_{b}\to \Lambda \omega)&=&1.6\times 10^{-6}
\end{eqnarray}
It is worth noticing that the experimental branching ratio for $%
\Lambda_{b}\to \Lambda J/\psi$ is $(4.7\pm 2.8)\times10^{-4}$. So our
calculated $\mathcal{BR}$ is within the experimental error. As for as the
branching ratio of $\Lambda_{b}\to \Lambda \rho^{0}$ and $\Lambda_{b}\to
\Lambda \omega$ are concerned, we have no experimental data for these
channels, so its hard to make any solid conclusion.

The numerical results for the helicity asymmetry parameter, $\alpha_{AS}$,
as defined in chapter 3, are summarized in Table [5.3], which can lead to a
complete determination of the polar angular distribution of the $\Lambda$
hyperon in the $\Lambda_{b}$ rest-frame.
\begin{table}[ht]
\caption{Helicity Asymmetry Parameter}\centering
\begin{tabular}{cccc}
\hline\hline
& $\Lambda_{b}\to \Lambda J/\psi$ & $\Lambda_{b}\to \Lambda \rho^{0}$ & $%
\Lambda_{b}\to \Lambda \omega$ \\ \hline\hline
$\alpha_{AS}$ & 46$\%$ & 43$\%$ & 43$\%$ \\ \hline\hline
\end{tabular}%
\end{table}

\subsection{Physical Observables for TR and CP violations}

As discussed in the previous chapter, we have several time reversal
violating parameters. Their numerical values corresponding to the decays $%
\Lambda_{b}\to \Lambda J/\psi$, $\Lambda_{b}\to \Lambda \rho^{0}$ and $%
\Lambda_{b}\to \Lambda \omega$ are summarized in Table [5.4]:
\begin{table}[ht]
\caption{Time reversal violating parameters}%
\begin{tabular}{ccccccc}
\hline\hline
V $(1^-)$ & $G_{T}^{\Lambda}$ & $\Delta G_{T}^{\Lambda}$ & $G_{TN}$ & $%
\Delta G_{TN}$ & $B_{T}$ & $B_{TN}$ \\ \hline\hline
$J/\psi$ & $2.9\times10^{-27} $ & $-1.1\times10^{-8} $ & $-4.9\times10^{-26}
$ & $-1.1\times10^{-26} $ & $1.0\times10^{-19} $ & $-1.7\times10^{-18} $ \\
$\rho^{0}$ & $-5.4\times10^{-27} $ & $-8.6\times10^{-10} $ & $%
-2.0\times10^{-29} $ & $5.5\times10^{-27}$ & $-4.1\times10^{-18} $ & $%
-1.5\times10^{-20} $ \\
$\omega$ & $-4.9\times10^{-27} $ & $-3.5\times10^{-10} $ & $%
5.2\times10^{-27} $ & $1.9\times10^{-27} $ & $-9.1\times10^{-18} $ & $%
9.6\times10^{-18} $ \\ \hline\hline
\end{tabular}%
\end{table}

The numerical values of CP-violating ratios, as defined in the previous
chapter, corresponding to the decays $\Lambda_{b}\to \Lambda J/\psi$, $%
\Lambda_{b}\to \Lambda \rho^{0}$ and $\Lambda_{b}\to \Lambda \omega$, are
summarized in Table [5.5] as:
\begin{table}[ht]
\caption{CP violating parameters}\centering
\begin{tabular}{cccc}
\hline\hline
CPV ratio & $J/\psi$ & $\rho^{0}$ & $\omega$ \\ \hline\hline
$R_{W}$ & $-0.03$ & $0.02$ & $0.62$ \\ \hline
$R_{L}$ & $7.02\times10^{-16}$ & $9.24\times10^{-16}$ & $4.99\times10^{-16}$
\\ \hline
$R_{T}$ & $-1.02$ & $1.05$ & $-0.48$ \\ \hline
$R_{N}$ & $0$ & $-1.99\times10^{-16}$ & $0$ \\ \hline
$R_{TT}$ & $-3.01\times10^{-16}$ & $-1.90\times10^{-16}$ & $0$ \\ \hline
$R_{TN}$ & $-9.36\times10^{-16}$ & $-8.30\times10^{-16}$ & $%
-4.52\times10^{-16}$ \\ \hline
$\gamma_{W}$ & $0.98$ & $-0.95$ & $2.06$ \\ \hline
$\gamma_{L}$ & $-0.99$ & $0.44$ & -2.27 \\ \hline
$\gamma_{T}$ & $-8.56\times10^{-17}$ & $-2.09\times10^{-16}$ & $%
-1.05\times10^{-16}$ \\ \hline
$\gamma_{N}$ & $14.23$ & -0.99 & $0.03$ \\ \hline
$\gamma_{TT}$ & $9.42\times10^{-17}$ & $7.59\times10^{-17}$ & $%
7.59\times10^{-17}$ \\ \hline
$\gamma_{TN}$ & $-0.70$ & $1.01$ & $-0.12$ \\ \hline\hline
\end{tabular}%
\end{table}

It interesting to note that non-zero value of the CP-violating ratios in
Table[5.5] are a clear signature of CP-violation in the baryonic sector. We
hope that in the forthcoming LHCb experiment these CP-asymmetric ratios will
be seen.

\section{Conclusions}

We have studied the decay process $\Lambda_{b}\to\Lambda V(1^{-})$ where we
considered the vector meson $V$ as either $J/\psi$, $\rho$ or $\omega$. In
our analysis, we have investigated the branching ratios $\mathcal{BR}$,
polarizations of the decay products and helicity symmetry violating
parameters for the same channels. We have also signaled out direct Time
Reversal $TRV$ and $CP$ violating observables in a model independent
analysis.

Thanks to the Jacob-Wick-Jackson helicity formalism, rigorous and detailed
calculations of the $\Lambda_{b}$-decays into one baryon and one vector
meson have been carried out completely. This helicity formalism allows us to
clearly separate the kinematical and dynamical contributions in the
computation of the amplitude corresponding to $\Lambda_{b}\to \Lambda
V(1^{-})$ decay. The cascade-type analysis is indeed very useful for
analysing polarization properties and Time Reversal effects since the
analysis of every decay in the decay chain can be performed in its
respective rest frame. In order to apply our formalism, all the numerical
calculations are done in $Mathematica-FeynCalc$. We also dealt at length
with the uncertainties coming from the input parameters. In particular,
these include the Cabibbo-Kobayashi-Maskawa matrix element parameters, $\rho$
and $\eta$, $etc$. Moreover, the heavy quark effective theory has been
applied in order to estimate the various form factors, $F_{i}(q2)$ and $%
G_{i}(q2)$, which usually describe dynamics of the electroweak transition
between two baryons. Corrections at the order of $O(1/mb)$ have been
included when the form factors were computed. In the calculation of b-baryon
decays, we need the Wilson coefficients, $C(mb)$, for the tree and penguin
operators at the scale $m_b$. One of the major uncertainties is that the
hadronic matrix elements for both tree and penguin operators involve
nonperturbative QCD. We have worked in the factorization approximation, with
$N_{c}=3$. Although one must have some doubts about factorization, it has
been pointed out that it may be quite reliable in energetic weak b-decays.

As regards theoretical results for the branching ratios $\Lambda_{b}\to
\Lambda J/\psi$, $\Lambda_{b}\to \Lambda \rho^{0}$ and $\Lambda_{b}\to
\Lambda \omega$, we made a comparison with PDG $\cite{pdg}$ for $%
\Lambda_{b}\to \Lambda J/\psi$ where an agreement is found as $\mathcal{BR}%
^{exp}(\Lambda_{b}\to \Lambda J/\psi) = 4.7 \pm 2.8\times 10^{-4}$ and our $%
\mathcal{BR}^{th}(\Lambda_{b}\to \Lambda J/\psi) = 6.3\times 10^{-4}$. So
this provide some justification for the theoretical calculations we made in
this thesis. For $\Lambda_{b}\to \Lambda \rho$ and $\Lambda_{b}\to \Lambda
\omega$, the lack of experimental results does not allow us to draw any
solid conclusions. However, we made their theoretical branching ratio
predictions, as $3.8\times10^{-6}$ and $1.6\times 10^{-6}$, respectively. In
this work we have not taken into account the final state interactions as
well as non-factorizable effects, which may have some important consequences
on these decays.

The determination of the helicity asymmetry parameter, $\alpha_{AS}$, for $%
\Lambda_{b}\to \Lambda J/\psi$, and $\Lambda_{b}\to \Lambda \rho(\omega)$,
has allowed us a complete determination of the polar angular distribution of
the $\Lambda$ hyperon in the $\Lambda_{b}$ rest-frame. In fact, the
knowledge of the $\Lambda$ polarization, $\mathcal{P}^{\Lambda}$, which
depends on the nature of the vector meson produced, in addition to that of
the SDM elements, $\rho_{ij}^{\Lambda_{b}}$, may be useful to calculate the
polar and azimuthal angular distributions of the proton (and pion) in the $%
\Lambda$ rest frame, resulting from decay $\Lambda\to p \pi$. In a similar
way, the polar and azimuthal angular distributions of leptons and
pseudo-scalar mesons in the vector meson rest-frame can also be computed.
From weak decays analysis, one knows that vector-polarizations of outgoing
resonances (or some of their components in appropriate frames) are T-odd
observables.

In our work, we have shown that some new observables can be measured: by
studying angular distributions of the transverse polarization vectors in the
decay planes of the intermediate resonances in the $\Lambda_{b}$ rest-frame.
We found that the magnitude of their effects is directly related to the $%
\Lambda_{b}$ polarization density matrix (PDM) and more precisely to the
non-diagonal elements, $\rho_{ij}^{\Lambda_{b}}$, appearing in the
interference terms of the decay amplitude.

Our Time reversal violating parameters, summarized in Table[5.4], have very
small values as compared to unity because we have used the Standard Model
for our numerical analysis. The values for these TR and CP violating
parameters may be enhanced in other models. These TR and CP violating
parameters may be experimentally observable at the forthcoming LHCb run.

We conclude this note with some remarks about the method considered for the
polarizations, angular distributions and C-P-T violating parameters.

\begin{enumerate}
\item Our analysis is completely model independent and is also independent
of spurious effects $\cite{TMA},\cite{datta1},\cite{BLSS}$ caused by final
state interactions $\cite{Wol}$, which which we have neglected in our
analysis. In particular, we stress that our tests for TRV do not rely on any
assumptions. Our calculation can be used as an input for calculating the
model predictions of the observables considered here $\cite{LA}$.

\item It is important to note that the TRV tests based on $\Lambda_{b}$
polarization are similar to those proposed for hyperon decays $\cite{gatto},%
\cite{Pakvasa}$. However in our case we may also consider the polarization
correlations $\cite{cw}$, which provide a TRV test independent of the
polarization of the parent resonance. Decays of the type $\Lambda_{b}\to
\Lambda V$ are very suitable for detecting possible TRV, as also pointed out
by other authors in studying CP violations $\cite{datta1},\cite{datta2}$.

\item The observables considered in the present work are very sensitive to
NP, since they are rid of unpleasant effects of Wilson's coefficients. These
quantities have been considered even more convenient than $B^{0} - \bar{B}%
^{0}$ mixing phases $\cite{datta1}$.

\item Reactions similar to those studied here have been proposed also by
other authors[41, 42] in a different context, for LHC forthcoming
experiment. Then it appears not unrealistic to suggest to measure also some
of the observables considered in the present note, that is, the angular
distribution and the polarization of at least one of the decay products.
\end{enumerate}

In our opinion, new fields of research like direct CP violation and T-odd
observables indicating a possible non-conservation of Time Reversal symmetry
can be investigated in the sector of beauty baryons and especially the $%
\Lambda_{b}$-bayons which can be copiously produced in the future hadronic
machine like LHC. In order to reach this aim, all uncertainties in our
calculations still have to be decreased, for example non-factorizable
effects have to be evaluated with more accuracy and final state effects
should be taken into consideration. Moreover, we strongly need more numerous
and accurate experimental data in $\Lambda_{b}$-decays, especially the $%
\Lambda_{b}$ polarization. We expect that our predictions should provide
useful guidance for future investigations and urge our experimental
colleagues to measure all the observables related to $\Lambda_{b}$ baryon
decays, if we want to understand direct CP violation and Time Reversal
symmetry better.

\section*{Acknowledgements}
I want to express my immense gratitude to my M.Phil supervisor, \textbf{Prof.
Riazuddin}, for his constant guidance, encouragement throughout this
thesis and for introducing me to the subject of Quantum Field Theory.
I am very thankful to \emph{Prof. Fayyazuddin} for many helpful discussions
that I had with him and teaching the best course on Particle Physics in
department of Physics. I am also thankful to \emph{Prof. Pervez Hoodbhoy}
who taught three valuable courses, which motivated me a lot to High Energy
Physics.

\end{document}